\documentclass[onecolumn]{aa}
\usepackage{graphicx}
\usepackage{wasysym}

\begin{document}

\title{Galactic tide and some properties of the Oort cloud}
\author{J. Kla\v{c}ka$^{1}$ \and L. K\'{o}mar$^{1}$ \and P. P\'{a}stor$^{1,2}$
\and M. Jur\v{c}i$^{1}$ \and E. H\"{o}nschov\'{a}$^{1}$}
\institute{Department of Astronomy, Physics of the Earth, and Meteorology \\
Faculty of Mathematics, Physics and Informatics, Comenius University \\
Mlynsk\'{a} dolina, 842 48 Bratislava, Slovak Republic \\
e-mails: klacka@fmph.uniba.sk, komar@fmph.uniba.sk \\
pavol.pastor@fmph.uniba.sk, jurci@fmph.uniba.sk
\and
Tekov Astronomical Observatory, \\
Sokolovsk\'{a} 21, 934~01, Levice, Slovak Republic}

\authorrunning{Kla\v{c}ka et al.}
\titlerunning{Properties of the Oort cloud}

\date{}

\abstract{
The paper deals with several properties of the Oort cloud of comets.
Sun, Galaxy (and Jupiter) gravitationally act on the comets.
New physical model of galactic tide is considered.
The main results can be summarized as follows: \\
1. Mass of the Oort cloud of comets is less than 1 mass of the Earth ($M_{E}$),
probably not greater than 1/2 $M_{E}$.  \\
2. Theoretical number of long-period comets with perihelion distance $q$ $<$ 5 AU is about 50-times
greater than the conventional approach yields. Gravity of Jupiter was taken into account in finding this result. \\
3. Semi-major axis $a$ and period of oscillations $P$ of eccentricity
(and other orbital elements) are related as $a^{3}$ $P$ $=$ 1 
in natural units for a moving Solar System in the Galaxy. 
The natural unit for time is the orbital period of the Solar System revolution around the galactic center
and the natural unit for measuring the semi-major axis is its maximum value for the half-radius of the
Solar System corresponding to the half-radius of the Oort cloud. 
The relation holds for the cases when comets approach the inner part of the Solar System,
e.g., perihelion distances are less than $\approx$ 100 AU. \\
4. The minimum value of semi-major axis for the Oort cloud is 
$a_{min}$ $\ll$ 1 $\times$ 10$^{4}$ AU. This condition was obtained both from the numerical 
results on cometary evolution under the action of the galactic tides and from the observational
distribution of long-period comets. If the density function of semi-major axis is approximated by 
proportionality $a^{\alpha}$, then $\alpha$ is $-$ 1/2, approximately. \\
5. The magnitude of the change in perihelion distance per orbit, $\Delta q$, of a comet due to galactic tides 
 is a strong function of semi-major axis $a$, proportional to $a^{8.25}$. 

}

\maketitle

\section{Introduction}
Intense modeling of the Oort cloud of comets started practically immediately after the Oort's paper (Oort 1950). 
As for the curent status of the ideas about the Oort cloud we refer to review papers, e. g., 
Dones et al. (2004), Levison and Jones (2007). 

New physical access to the theoretical modeling of the Oort cloud of comets 
was suggested by Kla\v{c}ka (2009a, 2009b). It is based on a new approach
to treating the effect of galactic tides, including improvement of
the model of the Galaxy. Its significance was discussed by
K\'{o}mar et al. (2009). Analytical approach to secular evolution
of orbital elements was discussed by P\'{a}stor et al. (2009).
The aim of this paper is to present some characteristics obtained from
the new model. The new model is represented by Eqs. (26)-(27) in Kla\v{c}ka (2009a) 
and denoted as the Model II in K\'{o}mar et al. (2009). The last paper
showed the importance of the new model in comparison with the conventional
models. Thus, we will not make any calculation based on the conventional models.
Instead of that we will concentrate on calculations based on the new modeling of galactic tides. 
The results of the calculations will be compared with the results obtained by other authors
isung conventional models of galactic tides.

Sec. 2 deals with distribution of cometary inclinations and it sheds some light
also on the value of the exponent $\alpha$ describing distribution of semi-major
axes of comets in the Oort cloud.
Sec. 3 shows that the magnitude of the change in perihelion
distance per orbit of a comet due to galactic tides is a strong function of
semi-major axis $a$, proportional to $a^{8.25}$.
Sec. 4 relates the semi-major axis of a comet and period of oscillation in other orbital elements
due to galactic tides. The found relation reminds the third Kepler's law, also in the case
when natural units of length and time are used. Secs. 5 and 6 discuss some theoretical
access to cometary distributions in eccentricity, perihelion distance and semi-major axis.
However, Sec. 6.3 presents also results obtained by numerical integration
of equation of motion for comets under the action of the Sun and the Galaxy.
Sec. 7 determines minimal perihelion distances $q_{min}$ and inclinations for the
instant when $q_{min}$ occurs. The section uses analytical approach to secular time
derivatives of orbital elements when the initial cometary inclinations
differ from 90$^{\circ}$. The two sources of gravity, the Sun and
the Galaxy, are used also in Sec. 8 dealing with the distribution
of comets in inclination to ecliptic. Sec. 9 improves the most
relevant previous results also for the case when the planet
Jupiter gravitationally influences the motion of the comets.
Finally, Sec. 10 deals with the distribution function in semi-major
axis and the number of comets in the Oort cloud, including estimate
of the mass of the Oort cloud.

\section{Distribution of cometary inclinations}
We are interested in distribution function of inclination of comets.
The inclination is the angle between a reference plane
and the instantaneous orbital plane of a comet.

Probability that a comet has an inclination lying in the interval
($i$ $-$ $di$/2, $i$ $+$ $di$/2) is
\begin{eqnarray}\label{eq:1}
dp &=& h(i) ~d \Omega ~,
\nonumber \\
d \Omega &=& 2 ~\pi ~ \sin i ~di ~,
\end{eqnarray}
where $d \Omega$ is the solid angle and $h(i)$ is a function,
$i \in \langle 0, \pi \rangle$. It is assumed that the normalization condition
\begin{eqnarray}\label{eq:2}
\int_{4 ~\pi} ~ h(i) ~d \Omega &=& 1~,
\nonumber \\
2 ~\pi ~ \int_{0}^{\pi} ~ h(i) ~ \sin i ~di  &=& 1
\end{eqnarray}
holds. The distribution function is
\begin{eqnarray}\label{eq:3}
H(i) &=& 2 ~ \pi ~ \int_{0}^{i} ~h(i') ~\sin i' ~di' ~,
\nonumber \\
H(\pi) &=& 1 ~.
\end{eqnarray}
The density function is 
\begin{eqnarray}\label{eq:4}
h_{H} (i) &\equiv& \frac{d H}{d i} = 2 ~ \pi ~ h(i)  ~\sin i ~.
\nonumber \\
H(\pi) &=& 1 ~.
\end{eqnarray}
The mean value is given by relation
\begin{equation}\label{eq:5}
\langle i \rangle =  2 ~ \pi ~ \int_{0}^{\pi} ~i' ~h(i') ~\sin i' ~di' ~,
\end{equation}

The special case of the isotropic distribution is defined by $h(i)$ $=$ constant.
Eqs. (\ref{eq:3})-(\ref{eq:5}) reduce to
\begin{eqnarray}\label{eq:6}
H(i) &=& \frac{1}{2} ~( 1 ~-~ \cos i ) ~,
\nonumber \\
\langle i \rangle &=& \frac{\pi}{2} ~,
\nonumber \\
h_{H}(i) &=& \frac{1}{2} ~ \sin i  ~.
\nonumber \\
h(i) &\equiv& \frac{1}{4 ~ \pi}  ~.
\end{eqnarray}
Equal numbers of prograde ($i$ $<$ $\pi$/2) and retrograde ($i$ $>$ $\pi$/2) orbits exist.

\subsection{Results of orbital evolution -- inclination to galactic equator}
We are dealing with distribution of comets in the Oort cloud in this subsection.
We are interested in the distribution of inclination measured with respect
to the galactic equatorial plane.

We made detailed numerical calculations of the orbital evolution of comets under the action of gravity 
of the Sun and Galaxy (Kla\v{c}ka 2009a, 2009b, K\'{o}mar et al. 2009). The results showed that initial inclination
of a comet decreased to some value if $i_{in}$ was smaller than 90$^{\circ}$
and increased to an another value if $i_{in}$ was greater than 90$^{\circ}$.
In reality the inclination changes on a time scale of millions of years.
Fig. 1 depicts such kind of evolution of inclination of a comet moving initially
on almost circular orbit in a distance of 5.0 $\times$ 10$^{4}$ AU.
We are interested in time average of the inclination in order to
find results comparable with statements presented in literature
(e.g., Duncan et al. 1987, Bailey 1983, Fern\'{a}ndez and Ip 1987,
Fern\'{a}ndez 1992, Fern\'{a}ndez and Gallardo 1999).

\begin{figure}[h]
\centering
\includegraphics[scale=0.65]{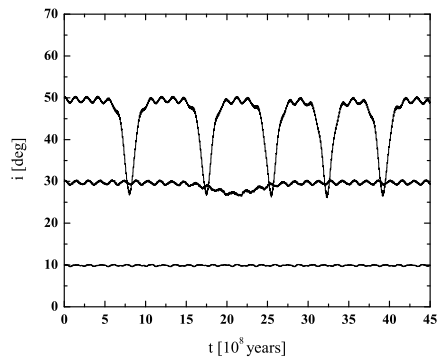}
\label{F1}
\caption{Time evolution of inclination of a comet moving initially
on almost circular orbit in a distance of 5.0 $\times$ 10$^{4}$ AU.
Three intial values of inclination are considered.}
\end{figure}

Table 1 presents the averaged time values of a comet with a given initial inclination.
On the basis of the data presented in Table 1 we can conclude
that the function $h(i)$ is not a constant. E.g, the interval
$i_{in}$ $\in$ $\langle 70^{\circ} , 90^{\circ} \rangle$ produces
the time averaged values
$i$ $\in$ ($60^{\circ} , 90^{\circ}$). An approximation yields
that for a mean value $i [ \mbox{deg} ]$ $\approx$ 60 $+$ (90$-$60)/2 $=$ 75
the following value holds:
$h$ (75$^{\circ}$) $\approx$ (90 $-$ 70)/(90 $-$ 60)/ (4 $\pi$) $=$
2 / 3 / (4 $\pi$). Similarly, the interval
$i_{in}$ $\in$ ($25^{\circ} , 70^{\circ}$) produces
the time averaged values
$i$ $\in$ ($25^{\circ} , 60^{\circ}$). An approximation yields
that for a mean value $i [ \mbox{deg} ]$ $\approx$ 25 $+$ (60$-$25)/2 $=$ 42.5
the following value holds:
$h$ (42.5$^{\circ}$) $\approx$ (70 $-$ 25)/(60 $-$ 25)/ (4 $\pi$) $\approx$
4/3 / (4 $\pi$). Thus, we can conclude that
$h ( \pi / 4 )$ $\approx$ (4 / 3) (4 $\pi$ )$^{-1}$.
One can see the difference between the value and the result presented in Eqs. (\ref{eq:6}).
The result $h ( \pi / 4 )$ $\approx$ (4 / 3) (4 $\pi$ )$^{-1}$ 
holds for semi-major axis $a$ $=$ 5.0 $\times$ 10$^{4}$ AU. 
The smaller $a$, the closer the result to that given in Eq. (\ref{eq:6}).

\begin{table}[h]
\centering
\begin{tabular}{|r|r|r|r|}
\hline
$i_{in}$ & $i$ & $i_{in}$ & $i$   \\
\hline
[deg] & [deg] & [deg] & [deg]  \\
\hline
\hline
  0.00 &   0.00 &  90.00 &  89.47 \\
 10.00 &   9.89 & 110.00 & 119.72 \\
 30.00 &  29.31 & 130.00 & 134.88 \\
 50.00 &  45.12 & 150.00 & 150.69 \\
 70.00 &  60.28 & 180.00 & 180.00 \\
\hline
\end{tabular}
\caption{Values of time averaged inclination $i$ for a comet with a given
initial inclination $i_{in}$ for initially almost circular orbit and
semi-major axis $a$ $=$ 5.0 $\times$ 10$^{4}$ AU due to the galactic tide.}
\label{tab:1}
\end{table}

Detailed numerical calculations show that the function $h$ defined by Eqs. (\ref{eq:1}) 
depends not only on the inclination $i$, but also on the semi-major axis $a$. The most relevant
results are presented in Table 2. The function for $a$ $=$
2.5 $\times$ 10$^{4}$ AU is depicted in Fig. 2, where also
isotropic function is shown, for comparison.

\begin{table}[h]
\centering
\begin{tabular}{|r|r|r|}
\hline
$a$ & $i_{max}$ & $h( i_{max}, a )$ / $h ( i = 0, a )$	 \\
\hline
[10$^{4}$ AU ] & [rad] & --- \\
\hline
\hline
$\rightarrow$ 0.0 & $\langle$ 0, $\pi / 2$ $\rangle$ & 1.0 \\
2.5 & 0.71 $\times$ $\pi / 4$ & $1.5$ \\
5.0 & 0.87 $\times$ $\pi / 4$ & $1.5$ \\
\hline
\end{tabular}
\caption{Maxima of the function $h (i, a)$,
$h_{max}$ $=$ $h( i_{max}, a )$, for various values of cometary
semi-major axes $a$. The maximum of $h$ holds for the angle $i_{max}$.
The interval $i$ $\in$ $\langle$ 0, $\pi / 2$ $\rangle$ is considered.}
\label{tab:2}
\end{table}

\begin{figure}[h]
\centering
\includegraphics[scale=0.28]{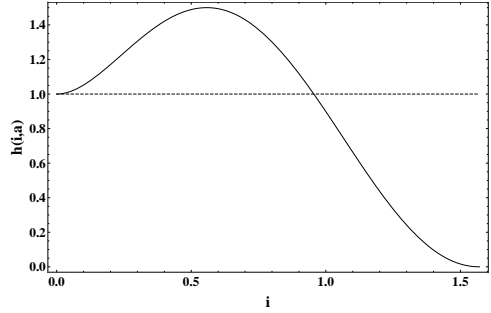}
\label{F2}
\caption{Function of inclination with respect to the galactic equatorial plane.
The solid line holds for the gravity of the Galaxy and the Sun and for the semi-major axis 
$a$ $=$ 2.5 $\times$ 10$^{4}$ AU.
The dotted line corresponds to the two-body problem.}
\end{figure}

The function can be approximated by the following conditions:
\begin{eqnarray}\label{eq:7}
\lim_{i \rightarrow 0} \frac{\partial h \left ( i, a \right ) }{\partial i} &=& 0 ~,
\nonumber \\
h \left ( \frac{\pi}{2}, a \right )  &=& 0 ~,
\nonumber \\
\lim_{i \rightarrow \pi / 2} \frac{\partial h \left ( i, a \right ) }{\partial i} &=& 0 ~,
\nonumber \\
i &=& i_{max} ~: ~~h ( i_{max}, a )  =
\max \left \{ h(i, a), i \in \langle 0, \frac{\pi}{2} \rangle \right \} ~,
\nonumber \\
\lim_{i \rightarrow i_{max}} \frac{\partial h \left ( i, a \right ) }{\partial i} &=& 0 ~.
\end{eqnarray}
Moreover,
\begin{eqnarray}\label{eq:8}
h \left ( \frac{\pi}{2} ~+~ i, a \right ) &=&
h \left ( \frac{\pi}{2} ~-~ i, a \right ) ~,
\nonumber \\
i &\in& \langle 0, \frac{\pi}{2} \rangle ~.
\end{eqnarray}

On the basis of the values presented in Table 2 and Eqs. (\ref{eq:7})
we can make the following approximation
\begin{eqnarray}\label{eq:9}
h (  i, a ) &=& h_{a} (i) ~\times ~ (1 + i + c_{12} ~ a ~ i^2 + c_{22} ~ a^2 ~ i^2 + c_{13} ~ a ~ i^3 + c_{23} ~ a^2 ~ i^3) ~,
\nonumber \\
h_{a} (i) &=& \left ( 2 - \frac{\pi}{2} \right ) ~ \left ( \frac{2}{\pi}  \right )^{3} ~ i^{3}
~+~ \left ( \pi - 3  \right )~  \left ( \frac{2}{\pi} \right )^{2} ~  i^{2} ~-~ i ~+~ 1  ~,
\nonumber \\
c_{12} &=& +~4.89184 ~,
\nonumber\\
c_{22} &=& -~0.80376 ~,
\nonumber\\
c_{13} &=& -~3.23854 ~,
\nonumber\\
c_{23} &=& +~0.67758 ~.
\end{eqnarray}
The function independent of semi-major axis is
\begin{eqnarray}\label{eq:10}
h(i) &\propto& \frac{\int\limits_{a_{min}}^{a_{max}}\,a^{\alpha}h(i,a)da}{\int\limits_{a_{min}}^{a_{max}}\,a^{\alpha}da} ~, 
\nonumber \\
\alpha &\in& ( -~ 4, -~ 2 ) ~; ~-~ 3/2 ~; ... ~.
\end{eqnarray}
We inserted also the values of the exponent $\alpha$ taken from Fern\'{a}ndez and Gallardo (1999), Duncan et al. (1987). 
We finally obtain
\begin{eqnarray}\label{eq:11}
h(i) &=& K  h_{a}(i) \left[ 1 + i - \frac{\ln\left(\frac{a_{max}}{a_{min}}\right)}{a_{max}^{-1}-a_{min}^{-1}} C_{1} (i) -
\frac{a_{max}-a_{min}}{a_{max}^{-1}-a_{min}^{-1}} C_{2} (i) \right] ~, ~~ \alpha = - 2 ~,
\nonumber \\
h(i) &=& K h_{a}(i) \left[ 1 + i + 2 \frac{a_{max}^{-1}-a_{min}^{-1}}{a_{max}^{-2}-a_{min}^{-2}} C_{1} (i) - 2
\frac{\ln\left(\frac{a_{max}}{a_{min}}\right)}{a_{max}^{-2}-a_{min}^{-2}} C_{2} (i) \right] ~, ~~ \alpha = - 3 ~,
\nonumber \\
h(i) &=& K h_{a}(i) \left [ 1 + i +\frac{\alpha+1}{\alpha+2} \frac{a_{max}^{\alpha+2}-a_{min}^{\alpha+2}}{a_{max}^{\alpha+1}-a_{min}^{\alpha+1}} C_{1} (i)
+ \frac{\alpha+1}{\alpha+3} \frac{a_{max}^{\alpha+3}-a_{min}^{\alpha+3}}{a_{max}^{\alpha+1}-a_{min}^{\alpha+1}} C_{2} (i) \right ] ~,
\nonumber \\
\alpha &\in& (- 4, - 3) \cup (- 3, - 2)  ~; ~-~ 3/2 ~; ... ~,
\nonumber \\
C_{1} (i) &=& c_{12} ~i^2 ~+~ c_{13} ~i^3 ~,
\nonumber \\
C_{2} (i) &=& c_{22} ~i^2 ~+~ c_{23} ~i^3 ~,
\end{eqnarray}
The quantity $K\equiv K ( \alpha, a_{min}, a_{max} )$ is determined by the
normalization condition
\begin{eqnarray}\label{eq:12}
2 ~\pi ~ \int_{0}^{\pi} ~ h(i, \alpha, a_{min}, a_{max}) ~ \sin i ~di  &=& 1 ~.
\end{eqnarray}
Having the function defined by Eqs. (\ref{eq:11})-(\ref{eq:12}), we can
find the distribution function defined by Eq. (\ref{eq:3}) and the observational
mean value $\langle i \rangle$. It follows
\begin{eqnarray}\label{eq:13}
\langle i \rangle &=& \frac{\pi}{2} ~.
\end{eqnarray}

Our detailed numerical calculations considering tidal effect of Galaxy show that
amplitude of semi-major axis $a$ rapidly increases with the value of $a$ when
$a$ $>$ 5 $\times$ 10$^{4}$ AU. If $a$ $=$ 7.5 $\times$ 10$^{4}$ AU, then
the amplitude of oscillation is 5 $\times$ 10$^{3}$ AU. Moreover, if
$a$ is comparable with 1 $\times$ 10$^{5}$ AU, then an increase of
semi-major axis exists. More correctly, there does not exist a constant value of
semi-major axis for secular evolution, an amplitude is about 25 $\times$ 10$^{3}$ AU.
Thus, the effect of galactic tide causes that stability of the Oort cloud exists only for 
$a$ $<$ $a_{max}$ and $a_{max}$ $\approx$ 0.8 $\times$ 10$^{5}$ AU. 
This seems to be consistent with observations, according to which
$a_{max}$ $\approx$ 1 $\times$ 10$^{5}$ AU and $a_{min}$ $\approx$ 1 
$\times$ 10$^{4}$ AU (see also Fig. 1 in Fern\'{a}ndez 1992), although the value
$a_{min}$ $=$ 1.5 $\times$ 10$^{4}$ AU is also presented
(Fern\'{a}ndez and Ip 1987).

Fig. 3 depicts the function(s) $h(i)$. Although the presented curves
do not exhibit any significant difference, the real results are relevant.
At first, if we take into account $a_{min}$ $\rightarrow$ 0, more correctly
$a_{min}$ $\ll$ 1 $\times$ 10$^{4}$ AU, then various values of $\alpha$
can be used and the function practically does not depend on the real value
of $\alpha$. The value $a_{min}$ $\rightarrow$ 0 is consistent with the data obtained 
from observational data on long-period comets (see Sec. 10), but it is not consistent with 
the conventional statements discussed in the previous paragraph.
However, if we would like to use the values not fulfilling $a_{min}$ $\rightarrow$ 0,
e.g., $a_{min}$ $=$ 1.0 $\times$ 10$^{4}$ AU or
$a_{min}$ $=$ 2.0 $\times$ 10$^{4}$ AU, then the only acceptable value of
$\alpha$ is $-$ 1: the value $\alpha$ $\ne$ $-$ 1 would produce nonpositive function $h$. 
The value $\alpha$ $=$ $-$ 1 is not consistent with the conventional 
statements on the Oort cloud that $\alpha$ $=$ $-$ 3/2 or $\alpha$ $\in$ ( $-$ 4, $-$ 2 )
(Duncan et al. 1987, Bailey 1983, Fern\'{a}ndez and Ip 1987,
Fern\'{a}ndez 1992, Fern\'{a}ndez and Gallardo 1999). Moreover, the observational 
data suggest that $\alpha$ is closer to zero: $\alpha$ $=$ 0.13 (see Eqs. \ref{eq:nca1-j} in Sec. 10),
or $\alpha$ $=$ $-$ 0.55 (see Eqs. \ref{eq:nca2-jn} in Sec. 10). 

\begin{figure}[h]
\centering
\includegraphics[scale=0.28]{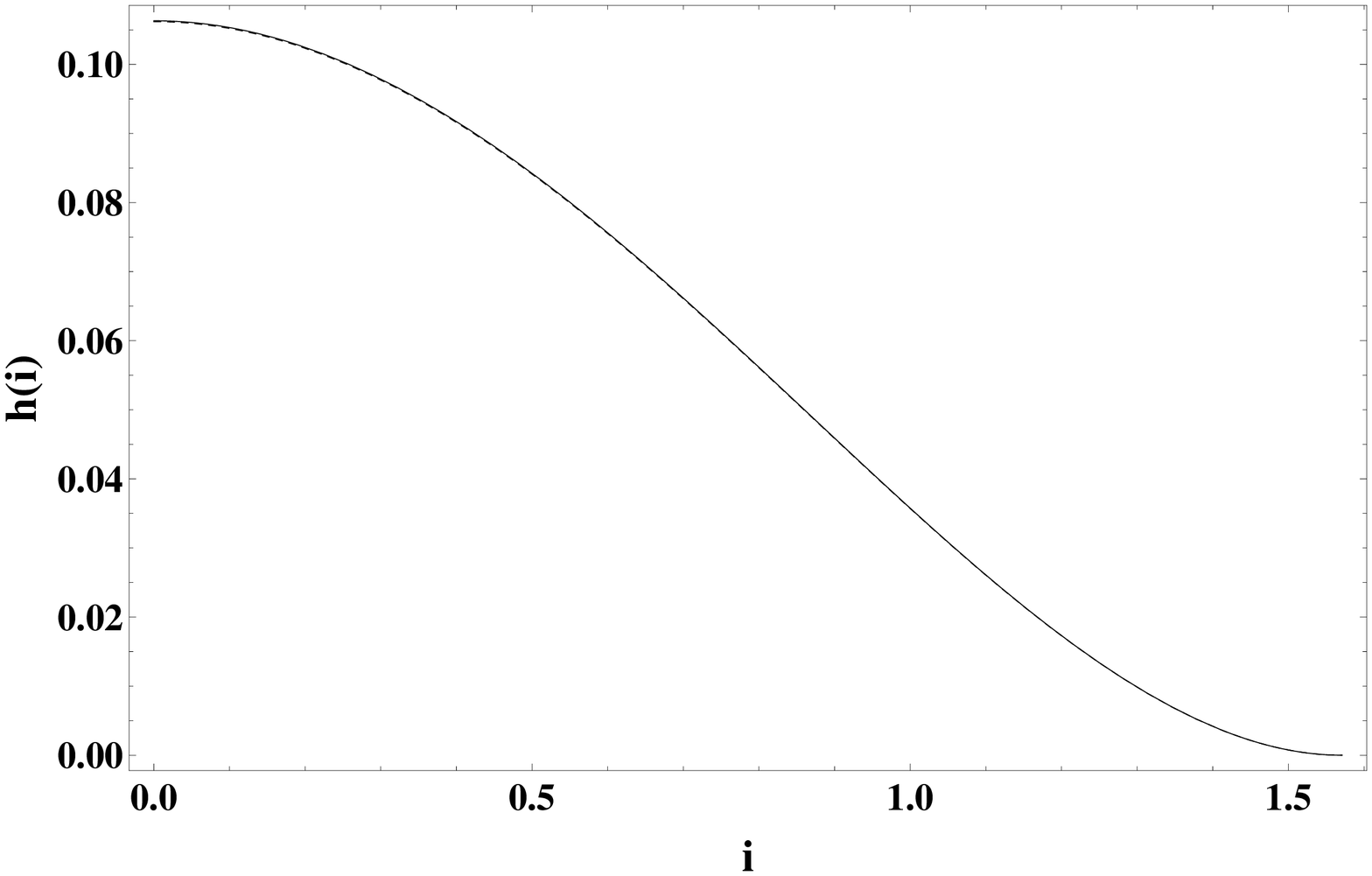}
\includegraphics[scale=0.28]{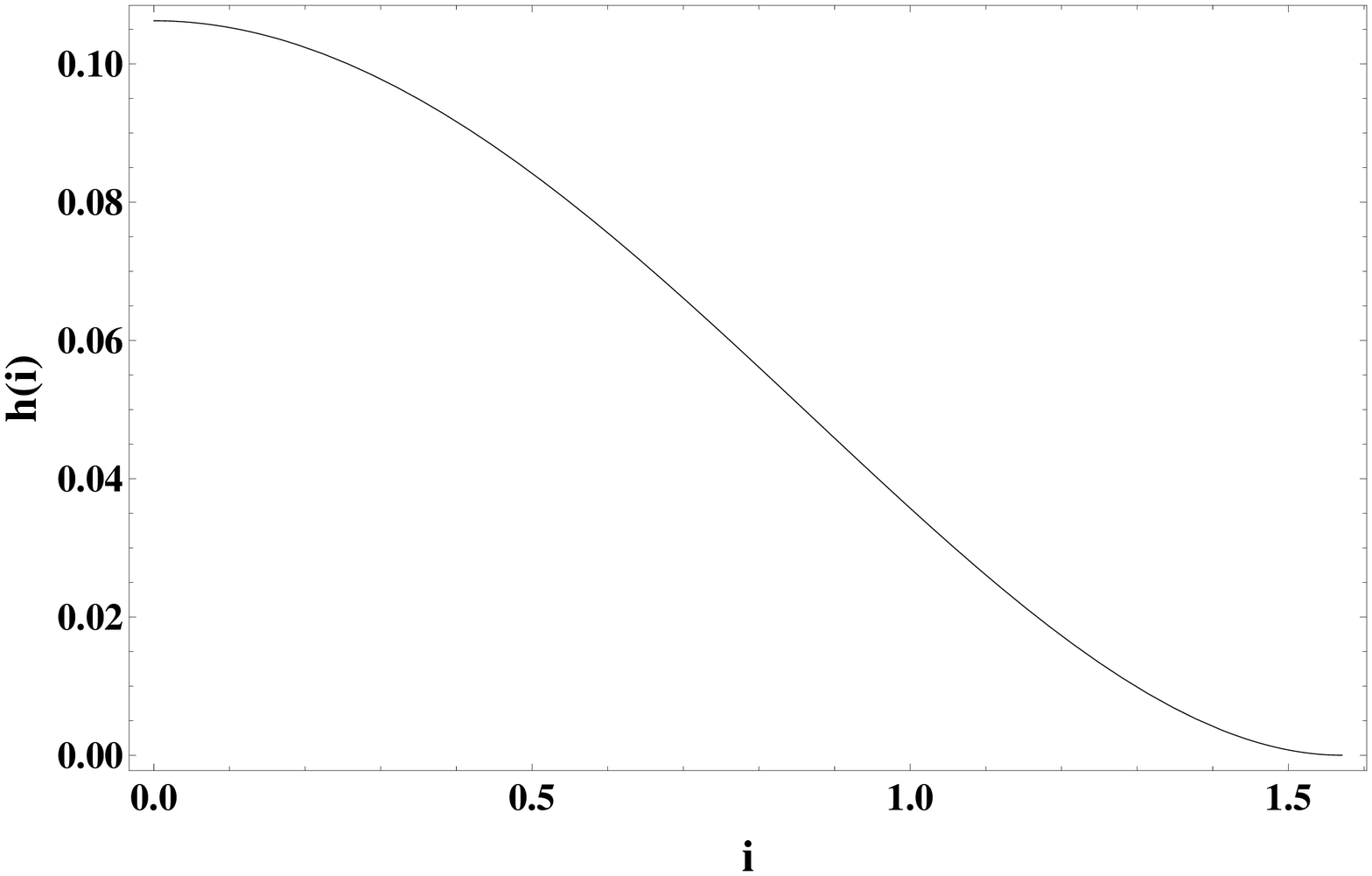}
\label{F3}
\caption{The function of inclination with respect to the galactic equatorial plane $h(i)$.
Averaging over distribution in semi-major axis is done.
The upper/left part of the figure is characterized by the values $a_{min}$ $\rightarrow$ 0,
$a_{max}$ $=$ 10 $\times$ 10$^{4}$ AU (practically independent of $\alpha$) .
The lower/right part of the figure corresponds to the case 
$a_{min}$ $=$ 1 $\times$ 10$^{4}$ AU, 
$a_{max}$ $=$ 10 $\times$ 10$^{4}$ AU and $\alpha$ $=$ $-$ 1.0.}
\end{figure}

\section{Timescales on which a comet's perihelion changes}
A formula for a timescales on which a comet's perihelion changes is presented
by Levison and Dones (2007, p. 583). However, as it is discussed by
K\'{o}mar et al. (2009), the formula does not correspond to reality.
It is important to find mathematical relation(s) which are based on physical
approach.

In order to obtain the cases when comets approach the inner part of the Solar System,
e.g., perihelion distances are less than $\approx$ 100 AU, we consider initial inclinations with respect
to the galactic equatorial plane about 90 degrees.

\subsection{First approach}
Let us consider the change of perihelion distance per revolution of a comet
around the Sun. The period of revolution of the comet is $T$ [years], the
change of perihelion distance during the period $T$ is $\Delta q$ [AU] and
semi-major axis of the comet is $a$ [AU]. Fig. 4 holds for the cases when $q$ obtains its minimum
values during the time evolution $q(t)$ for 4.5 $\times$ 10$^{9}$ years.
The solid line is an analytical approximation to numerical results obtained
from evolution for the model of Kla\v{c}ka (2009a, Eqs. 26-27; Model II in K\'{o}mar et al. 2009).
We can conclude that
\begin{eqnarray}\label{eq:LDperi1}
\log_{10} \left ( \frac{\Delta q}{T} \right ) &=& A_{1} ~+~ A_{2} ~ \log_{10} a~,
\nonumber \\
A_{1} &=& -~35.95 \pm 0.89 ~,
\nonumber \\
A_{2} &=& +~6.75 \pm 0.20 ~.
\end{eqnarray}

\begin{figure}[h]
\centering
\includegraphics[scale=0.65]{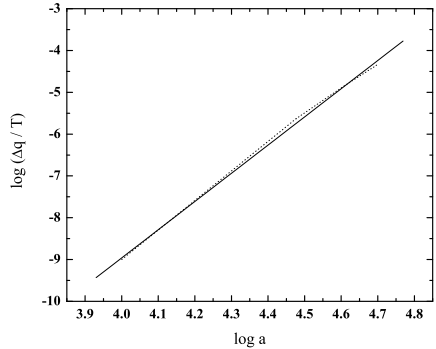}
\label{F4}
\caption{Change of perihelion distance $\Delta q$ [AU] per revolution of a comet
around the Sun. The period of revolution of the comet is $T$ [years] and
semi-major axis of the comet is $a$ [AU]. Solid line corresponds to linear fit
of the numerical solution (dashed curve) for the tidal effect of the Galaxy.}
\end{figure}

\subsection{Second approach}
We want to find a formula which is more similar to the mathematical formula
presented by Levison and Dones (2007, p. 583). As it was already stressed,
our formula must respect physical reality. We can use detailed numerical
calculation for orbital evolution for our new model (Kla\v{c}ka 2009a, 2009b;
Model II in K\'{o}mar et al. 2009), or we can
try to use the analytical approach to secular evolution of orbital elements
(discussed by P\'{a}stor et al. 2009). The analytical approach
is limited by the condition that semi-major axis of a comet must be less
than about 1.5 $\times$ 10$^{4}$ AU if the condition
$T$ $\times$ ( 1 / $T_{z}$ $+$ 1 / $T_{0}$) $<$ 0.05, where
$T_{z}$ is the period of oscillations of the Sun with respect to the
galactic equatorial plane and $T_{0}$ is the period of revolution of the
Sun with respect to the center of the Galaxy.

Detailed solution of the equation of motion corresponding to Model II yields
\begin{eqnarray}\label{eq:LDperi2}
\log_{10} \left ( \frac{\Delta q}{T} \right ) &=& A_{1} ~+~ A_{2} ~ \log_{10} a ~+~ A_{3} ~ \log_{10} q~,
\nonumber \\
A_{1} &=& -~33.67 \pm 1.48 ~,
\nonumber \\
A_{2} &=& +~6.29 \pm 0.31 ~,
\nonumber \\
A_{3} &=& -~0.12 \pm 0.07 ~.
\end{eqnarray}
Again, as in the previous subsection, the period of revolution of the comet is $T$ [years], the
change of perihelion distance during the period $T$ is $\Delta q$ [AU] and
semi-major axis and perihelion distance of the comet are $a$ [AU] and $q$ [AU].
Eqs. (\ref{eq:LDperi2}) contain the same quantities as the relation presented by Levison and Dones (2007, p. 583).
It is immediately seen that the values of $A_{2}$ and $A_{3}$ significantly differ from the values
presented by Levison and Dones.

\subsection{Comparison of the approaches}
We have found two relations, one of them is represented by Eqs. (\ref{eq:LDperi1}), the other one by Eqs. (\ref{eq:LDperi2}).
None of them is consistent with the relation presented by Levison and Dones (2007, p. 583).
Moreover, the relation represented by Eqs. (\ref{eq:LDperi2}) is characterized by much greater errors of the
coefficients than it is in the case of Eqs. (\ref{eq:LDperi1}).

Secular approach to the orbital evolution treated by P\'{a}stor et al. (2009) also confirms that
the relation presented by Levison and Dones (2007, p. 583) is not consistent with physics of the
galactic tide.

According to Eqs. (\ref{eq:LDperi1})-(\ref{eq:LDperi2}) we see that the magnitude of the change in
perihelion distance per orbit, $\Delta q$, of a comet due to galactic tides is a strong function of semi-major axis
$a$, proportional to $a^{8.25}$, or considering some error, $a^{\delta}$, $\delta$ $\in$ (7.5, 8.5). 
This is a new result, not consistent with the conventional value $\delta$ $=$ 3.5 (see, e.g., Dones et al. 2004, p. 155). 

\section{Relation between semi-major axis and oscillation period}
Galactic tide causes oscillations of eccentricity (perihelion and aphelion distances, angular orbital elements) of a comet in the Oort cloud
(see Fig. 7 in K\'{o}mar et al. 2009). We are interested in a relation between the semi-major axis $a$ and the oscillation
period $P$ of the comet. 

We are interested in the cases when comets approach the inner part of the Solar System,
e.g., perihelion distances are less than $\approx$ 100 AU. Thus, we consider the values of about 90 degrees
for the initial inclinations with respect to the galactic equatorial plane. Various values of other orbital elements are taken into account.

Detailed numerical calculations for 1 $\times$ 10$^{4}$ AU $<$ $a$ $<$ 8 $\times$ 10$^{4}$ AU
yield the result presented in Fig. 5, which yields
\begin{eqnarray}\label{eq:apnew}
\log_{10} a  &=&  A~\log_{10} P  ~+~ B ~,
\nonumber \\
A &=& -~0.362  \pm 0.029 ~,
\nonumber \\
B &=& +~7.846 \pm 0.254 ~,
\nonumber \\
\left [ a \right  ] &=&  \mbox{AU} ~, ~~ \left [ P  \right ] = \mbox{yr} ~.
\end{eqnarray}

\begin{figure}[h]
\centering
\includegraphics[scale=0.65]{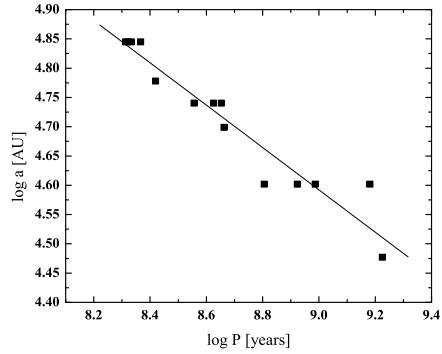}
\label{F5}
\caption{Semi-major axis $a$ as a function of oscillation period $P$ of eccentricity
of a comet from the Oort cloud. The oscillations are caused by the effect of galactic tide.
Solid line corresponds to linear fit of the data obtained by numerical solution for the tidal
effect of the Galaxy. The data for initial inclination 90 degrees are used (other values of
inclinations may yield a little different values).}
\end{figure}

Since the period of oscillations slightly changes on the scales of billions of years, we can rewrite Eq. (\ref{eq:apnew}) into a more
simple relation $a ^{3} ~P$ $=$ $constant$. The value of the $constant$ is determined by the least square method. We obtain
\begin{eqnarray}\label{eq:apnewf}
a ^{3} ~P &=& constant ~,
\nonumber \\
constant  &=& ( 6.93  \pm 0.23 ) \times 10^{22} \mbox{AU}^{3}~\mbox{yr} ~,
\nonumber \\
\left [ a \right  ] &=&  \mbox{AU} ~, ~~ \left [ P  \right ] = \mbox{yr} ~.
\end{eqnarray}

In natural units for the moving Solar System in the Galaxy we obtain
\begin{eqnarray}\label{eq:apnewf1}
a ^{3} ~P &=& 1 ~.
\end{eqnarray}
The natural unit for time is the orbital period of Solar System revolution around the galactic center,
$2 \pi / (A - B)$ $=$ 2.3 $\times$ 10$^{8}$ yrs.
The natural unit for the semi-major axis is its maximum value for the half-radius of the
Solar System which equals to the half-radius of the Oort cloud, 0.7 $\times$ 10$^{5}$ AU.
The value yields limiting cometary aphelion distances 1.4 $\times$ 10$^{5}$ AU, half of
the distance between the Sun and its nearest star.

As a first approximation, a comet of the Oort cloud revolves around the Sun on an ellipse fulfilling
the third Kepler's law $a^{3} / T^{2}$ $=$ $constant (two~body~problem)$, where $T$ is period of revolution
of the comet. Moreover, eccentricity of the comet oscillates due to the effect of galactic tide and
the period of oscillations of the eccentricity relates the semi-major axis of the comet as
$a^{3} ~P$ $=$ $constant (galactic~tide)$.

\section{Probability that eccentricity is less than a given value}
We are interested in the probability $P( e < e_{L} )$ that eccentricity $e$ of comets is less than
a given value $e_{L}$.

Orbital evolution of eccentricity due to the effect of galactic tides is a periodic
function of time, approximately. We will use the following approximation:
\begin{eqnarray}\label{eq:ecc}
e ( t ) &=& \frac{1}{2} ~ \left \{ 1 ~-~ \sin \left [ \frac{\pi}{2} \left (  4~ \frac{t}{ P} ~-~ \delta_{in} \right ) \right ] \right \} ~,
\nonumber \\
e ( t = 0) &\equiv& e_{in} = \frac{1}{2} ~ \left [ 1 ~+~ \sin \left ( \frac{\pi}{2} ~ \delta_{in} \right ) \right ]  ~,
\nonumber \\
\delta_{in} &=& \frac{2}{\pi} ~ \arcsin \left ( 2 e_{in} ~-~ 1 \right ) ~,
\nonumber \\
\delta_{in} &\in& \langle -~1, + 1 \rangle ~,~~~ e_{in} \in \langle 0, +1 \rangle ~,
\end{eqnarray}
for the time evolution of eccentricity $e(t)$ with the period of oscillations $P$,
as it is discussed in the previous section.

\subsection{Density and distribution functions of eccentricity}
Let an initial distribution of eccentricities of comets in the Oort cloud is represented by a density function
$f(e_{in})$. We are interested in the density function $f(e)$ at a given time $t$.
We have $e$ $=$ $e(t)$, $e(t=0)$ $=$ $e_{in}$. Gravitation of the Sun and Galaxy will be considered.

If $F$ is the distribution function, then
\begin{eqnarray}\label{eq:eccdF}
dF &=&  f[e (t)] ~ de ~,
\nonumber \\
dF &=& f(e_{in}) ~ de_{in} ~.
\end{eqnarray}

Knowing the inverse function $e_{in}=g^{-1}(e)$ to the function $g(e_{in})$, Eqs. (\ref{eq:eccdF}) yield
\begin{eqnarray}\label{eq:eccfe}
f(e)=f[g^{-1}(e)] ~\frac{1}{ \left | g'[g^{-1}(e)] \right | } ~.
\end{eqnarray}

Let us rewrite Eq. (\ref{eq:ecc}) into the form
\begin{eqnarray}
e=g (e_{in}) = \frac{1}{2}~ \left[1-2 \sin \left (\frac{2\pi t}{P}\right)\sqrt{e_{in}(1-e_{in})}+\cos\left(\frac{2\pi t}{P}\right)(2e_{in}-1)\right] ~.
\end{eqnarray}
Since this function is not monotonous, it's inverse function does not exist. However, it is possible to
split the function $g (e_{in})$ into two monotonous functions. There exist inverse functions for the two monotonous functions:
\begin{eqnarray}\label{eq:inv1}
g^{-1}(e) &=& -~ \sqrt{e(1-e)} \left| \sin\left(\frac{2\pi t}{P}\right)\right| + \frac{1}{2} \left[1+(2e-1)\cos\left(\frac{2\pi t}{P}\right)\right] ~,
\nonumber \\
e_{in} &\in& \langle 0, \sin^2 \left ( \frac{\pi t}{P}\right ) \rangle ~,
\end{eqnarray}
and
\begin{eqnarray}\label{eq:inv2}
g^{-1}(e) &=& +~ \sqrt{e(1-e)}\left|\sin\left(\frac{2\pi t}{P}\right)\right|+\frac{1}{2}\left[1+(2e-1)\cos\left(\frac{2\pi t}{P}\right)\right] ~,
\nonumber \\
e_{in} &\in& \langle \sin^2 \left ( \frac{\pi t}{P} \right), 1\rangle ~.
\end{eqnarray}

Eqs. (\ref{eq:eccfe}), (\ref{eq:inv1})-(\ref{eq:inv2}) yield
\begin{eqnarray}\label{eq:f1}
f (e) &=& f \left [ g^{-1} \left ( e \right ) \right ]
\frac{1}{\left | \cos \chi ~+~ X_{e1} \right |} ~,
\nonumber \\
X_{e1} &=& \frac{(2e-1) \cos \chi ~ \sin \chi  ~-~ 2 \sqrt{e(1-e)} \left | \sin \chi \right | \sin \chi}{
\sqrt{1 + \left [ 1 + 8 e (e-1) \right ] \sin^2 \chi - (2e-1)^2~+~ 4 ( 2 e - 1 ) \sqrt{e(1-e)} \left|\sin \chi \right|\cos\chi}} ~,
\nonumber \\
e_{in} &=& g^{-1}(e) = -~ \sqrt{e(1-e)} \left | \sin\chi \right | + \frac{1}{2} \left [ 1 + (2e-1) \cos \chi \right ] ~,
\nonumber \\
\chi &=& \frac{2~\pi}{P} ~t ~,
\nonumber \\
e_{in} &\in& \langle 0, \sin^2 \left ( \frac{\pi t}{P}\right ) \rangle ~,
\end{eqnarray}
and,
\begin{eqnarray}\label{eq:f2}
f(e) &=& f \left [ g^{-1} \left ( e \right ) \right ]
\frac{1}{ \left | \cos \chi ~+~ X_{e2} \right | } ~,
\nonumber \\
X_{e2} &=& \frac{(2e-1) \cos \chi ~ \sin \chi  ~+~ 2 \sqrt{e(1-e)} \sin^2 \chi}{\sqrt{1 + \left [ 1 + 8 e (e-1) \right ] \sin^2 \chi - (2e-1)^2 ~-~ 4 ( 2 e - 1 ) \sqrt{e(1-e)} \left|\sin \chi \right|\cos\chi}} ~,
\nonumber \\
e_{in} &=& g^{-1}(e) = +~ \sqrt{e(1-e)} \left | \sin \chi \right | + \frac{1}{2} \left [ 1 + (2e-1) \cos \chi \right ] ~,
\nonumber \\
\chi &=& \frac{2~\pi}{P} ~t ~,
\nonumber \\
e_{in} &\in& \langle \sin^2 \left ( \frac{\pi t}{P} \right), 1 \rangle ~.
\end{eqnarray}

The distribution function is
\begin{eqnarray}\label{eq:eccdf}
F(e) &=& \int_{0}^{e} ~f(e') ~de' ~,
\nonumber \\
F(1) &=& 1 ~.
\end{eqnarray}

We may introduce that instead of Eqs. (\ref{eq:eccdF}) an alternative approach can be used. Knowing the
density function $f(e_{in})$ and $e=g(e_{in})$, we may use a convolution for finding the density function $f(e)$.
The convolution yields
\begin{eqnarray}\label{eq:konvolucia}
f(e) &=& \int ~f(e_{in}) ~ \delta \left [ e ~-~ g( e_{in}) \right ] ~de_{in} ~.
\end{eqnarray}

\subsubsection{Uniform distribution}
Let us consider, as an example, a uniform distribution of eccentricities.
The density function is
\begin{eqnarray}\label{eq:ud1}
f(e_{in}) &=& 1~, ~~~ e_{in} \in \langle 0, 1 \rangle ~.
\end{eqnarray}
The distribution function is $F(e_{in})$ $=$ $\int_{0}^{e_{in}}$ $f (e_{in}')~ de_{in}'$ $=$ $e_{in}$, $F(1)$ $=$ 1.

Using Eqs. (\ref{eq:f1})-(\ref{eq:f2}), we obtain
\begin{eqnarray}\label{eq:udf1}
f (e) &=& \frac{1}{\left | \cos \chi ~+~ X_{e1} \right |} ~,
\nonumber \\
X_{e1} &=& \frac{(2e-1) \cos \chi ~ \sin \chi  ~-~ 2 \sqrt{e(1-e)} \left | \sin \chi \right | \sin \chi}{
\sqrt{1 + \left [ 1 + 8 e (e-1) \right ] \sin^2 \chi - (2e-1)^2~+~ 4 ( 2 e - 1 ) \sqrt{e(1-e)} \left|\sin \chi \right|\cos\chi}} ~,
\nonumber \\
\chi &=& \frac{2~\pi}{P} ~t ~,
\nonumber \\
e_{in} &\in& \langle 0, \sin^2 \left ( \frac{\pi t}{P}\right ) \rangle ~,
\end{eqnarray}
and,
\begin{eqnarray}\label{eq:udf2}
f(e) &=& f \left [ g^{-1} \left ( e \right ) \right ]
\frac{1}{ \left | \cos \chi ~+~ X_{e2} \right | } ~,
\nonumber \\
X_{e2} &=& \frac{(2e-1) \cos \chi ~ \sin \chi  ~+~ 2 \sqrt{e(1-e)} \sin^2 \chi}{\sqrt{1 + \left [ 1 + 8 e (e-1) \right ] \sin^2 \chi - (2e-1)^2 ~-~ 4 ( 2 e - 1 ) \sqrt{e(1-e)} \left|\sin \chi \right|\cos\chi}} ~,
\nonumber \\
\chi &=&  \frac{2~\pi}{P} ~t ~,
\nonumber \\
e_{in} &\in& \langle \sin^2 \left ( \frac{\pi t}{P} \right), 1 \rangle ~.
\end{eqnarray}

As an illustration we take $t$ $=$ $P /2$. Eqs. (\ref{eq:udf1})-(\ref{eq:udf2}) yield $f(e)$ $=$ 1, and, the distribution function is $F(e)$ $=$ $e$:
\begin{eqnarray}\label{eq:udf3}
f(e) &=& 1 ~, ~~ t = 0 ~,
\nonumber \\
F(e) &=& e ~, ~~ t = 0 ~,
\nonumber \\
f(e) &=& 1 ~, ~~ t = P/2 ~,
\nonumber \\
F(e) &=& e ~, ~~ t = P/2 ~.
\end{eqnarray}

\subsubsection{Distribution function $F(e_{in})$ $=$ $e_{in}^{2}$}
We will consider, as an another example, $F(e_{in})$ $=$ $e_{in}^{2}$. Motivation for this distribution function comes
from Fern\'{a}ndez and Gallardo (1999), Hills (1981) and Jeans (1919). The corresponding density function is
\begin{equation}\label{eq:jeans}
f(e_{in}) = 2~e_{in} ~,~~ e_{in} \in \langle 0, 1 \rangle ~.
\end{equation}

Using Eqs. (\ref{eq:f1})-(\ref{eq:f2}), we obtain
\begin{eqnarray}\label{eq:jeansf1}
f (e) &=& \frac{1+(2e-1) \cos \chi ~-~ 2 \sqrt{e(1-e)} \left |\sin \chi \right|}{\left | \cos \chi ~+~ X_{e1} \right |} ~,
\nonumber \\
X_{e1} &=& \frac{(2e-1) \cos \chi ~ \sin \chi  ~-~ 2 \sqrt{e(1-e)} \left | \sin \chi \right | \sin \chi}{
\sqrt{1 + \left [ 1 + 8 e (e-1) \right ] \sin^2 \chi - (2e-1)^2~+~ 4 ( 2 e - 1 ) \sqrt{e(1-e)} \left|\sin \chi \right|\cos\chi}} ~,
\nonumber \\
\chi &=& \frac{2~\pi}{P} ~t ~,
\nonumber \\
e_{in} &\in& \langle 0, \sin^2 \left ( \frac{\pi t}{P}\right ) \rangle ~,
\end{eqnarray}
and,
\begin{eqnarray}\label{eq:jeansf2}
f(e) &=& \frac{1+(2e-1) \cos \chi  ~+~ 2 \sqrt{e(1-e)} \left |\sin \chi \right|}{ \left | \cos \chi ~+~ X_{e2} \right | } ~,
\nonumber \\
X_{e2} &=& \frac{(2e-1) \cos \chi ~ \sin \chi  ~+~ 2 \sqrt{e(1-e)} \sin^2 \chi}{\sqrt{1 + \left [ 1 + 5 e (e-1) \right ]  
\left | \sin \chi \right | \sin \chi - (2e-1)^2 ~-~ 2 ( 2 e - 1 ) \sqrt{e(1-e)} \left|\sin \chi \right|\cos\chi}} ~,
\nonumber \\
\chi &=&  \frac{2~\pi}{P} ~t ~,
\nonumber \\
e_{in} &\in& \langle \sin^2 \left ( \frac{\pi t}{P} \right), 1 \rangle ~.
\end{eqnarray}

As an illustration we take $t$ $=$ $P/2$. Eqs. (\ref{eq:jeansf1})-(\ref{eq:jeansf2}) yield $f(e)$ $=$ 2 ( 1 $-$ $e$ ), and, the distribution function is 
$F(e)$ $=$ $e$ ( 2 $-$ $e$ ):
\begin{eqnarray}\label{eq:jeansf3}
f(e) &=& e ~, ~~ t = 0 ~,
\nonumber \\
F(e) &=& \frac{1}{2} ~e^{2} ~, ~~ t = 0 ~,
\nonumber \\
f(e) &=& 2 ~( 1 ~-~ e ) ~, ~~ t = P/2 ~,
\nonumber \\
F(e) &=& e ~( 2 ~-~ e ) ~, ~~ t = P/2 ~.
\end{eqnarray}
The situation represented by Eqs. (\ref{eq:jeansf3}) is illustrated in Fig. 6. 

\begin{figure}[h]
\centering
\includegraphics[scale=0.65]{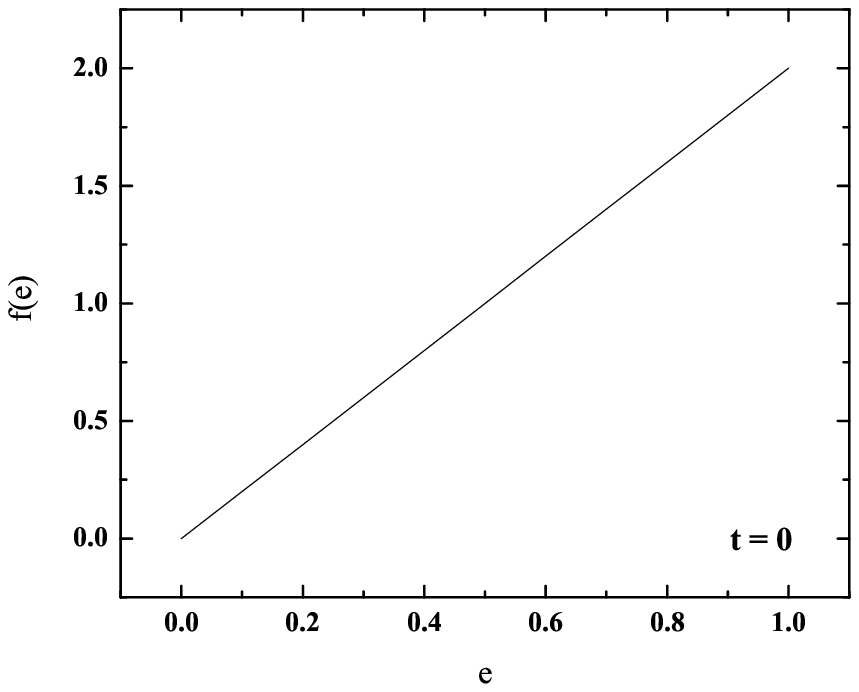}
\includegraphics[scale=0.65]{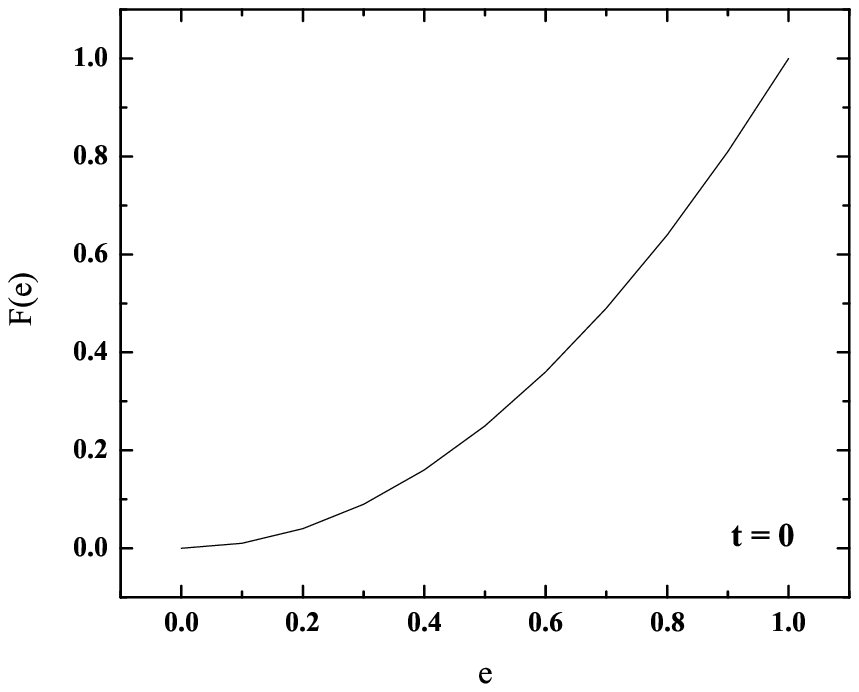}
\includegraphics[scale=0.65]{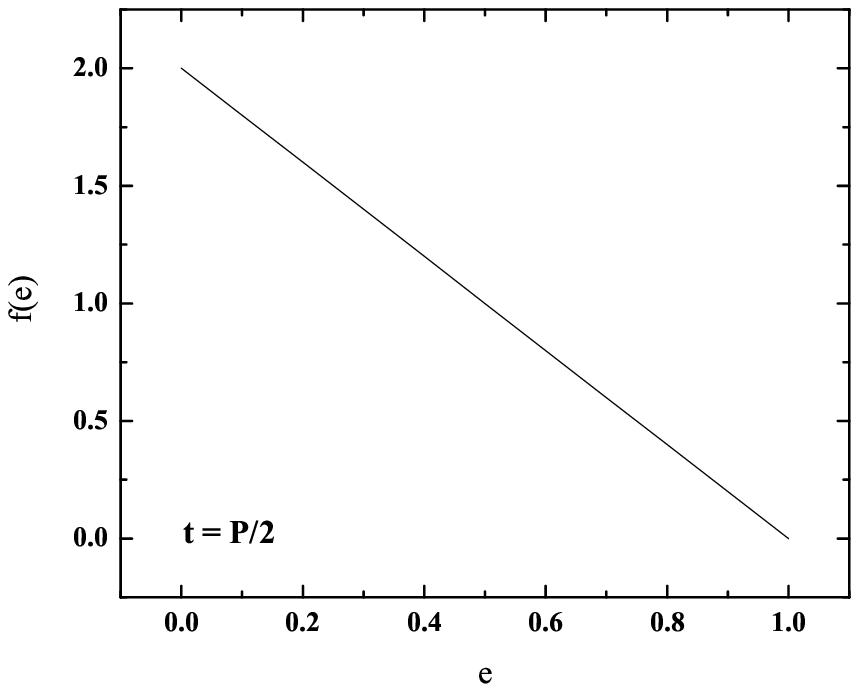}
\includegraphics[scale=0.65]{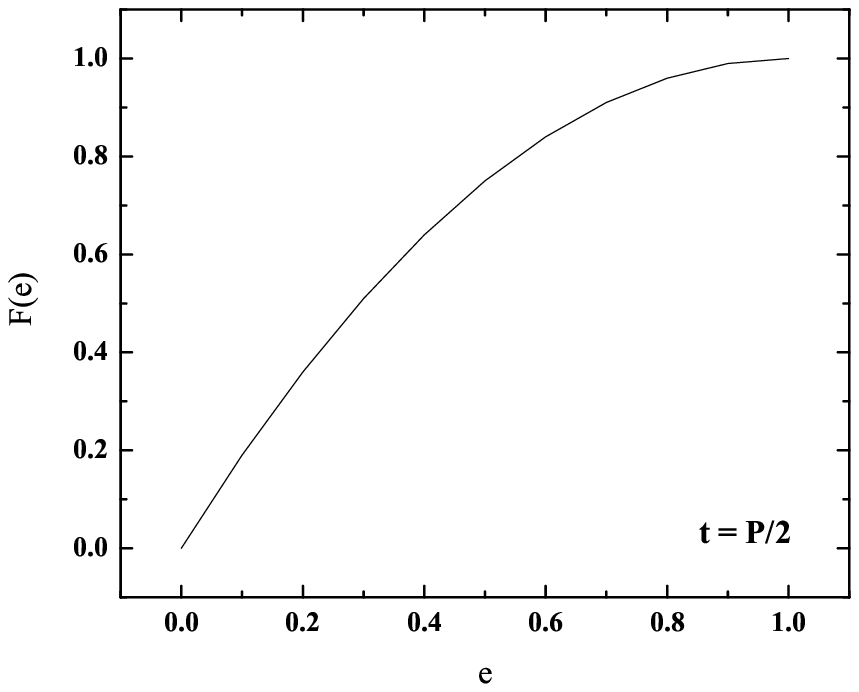}
\label{F6}
\caption{Density $f(e)$ and distribution $F(e)$ functions of eccentricity for two instants: 
$t$ $=$ 0 and $t$ $=$ $P/2$, where $P$ is the oscillation period of eccentricities 
(perihelion and aphelion distances, inclinations, ...).
The oscillations are caused by the effect of galactic tide.}
\end{figure}

\subsection{$P ( e < e_{L} )$ for a random time}
Now, we want to take into account some other important facts.
At first, we do not know at which time a given distribution $f(e_{in})$ holds.
Then, we know from the previous section that there is a relation between the semi-major axis $a$ and
period of oscillations $P$. Thus, comets with various values of $a$ are characterized
with various values of $P$. Moreover, we do not know the exact value of the parameter
$\alpha$ defining the distribution in semi-major axis, as it is presented in Sec. 2.
All these facts lead us to the conclusion that we have to find an another
possibility of obtaining the probability $P( e < e_{L} )$ that eccentricity of comets $e$ is less than
a given value $e_{L}$.

Let us consider Eqs. (\ref{eq:ecc}). We can write
\begin{eqnarray}\label{eq:peel}
P ( e < e_{L} ) &=& \frac{\Delta t_{L} }{P}  ~,
\nonumber \\
\Delta t_{L} &=& | t_{L~1} ~-~ t_{L~2} | ~,
 \nonumber \\
e_{L} &=& \frac{1}{2} ~
\left \{ 1 ~-~ \sin \left [ \frac{\pi}{2} \left (  4~ \frac{t_{L}}{ P} ~-~ \delta_{in} \right ) \right ] \right \} ~.
\end{eqnarray}
The last equation for $e_{L}$ determines the two values $t_{L~1}$ and $t_{L~2}$. Thus, we finally obtain
\begin{eqnarray}\label{eq:peelf}
P ( e < e_{L} ) &=& \frac{1}{2} ~-~ \frac{1}{\pi} ~\arcsin \left ( 1 ~-~ 2~e_{L} \right ) ~.
\end{eqnarray}

\section{Probability that perihelion distance is less than a given value}
We are interested in the probability $P( q < q_{L} )$ that perihelion distance $q$ of comets is less than
a given value $q_{L}$. We will assume that all comets have the same semi-major axis $a$.
Sec. 6.4 will consider joint/cumulative distribution function consisting of two marginal distributions, 
one in $a$ and another one in $q$.

Let the orbital evolution of eccentricity due to the effect of galactic tides is given by Eqs. (\ref{eq:ecc}).
Since $q$ $=$ $a$ ( 1 $-$ $e$ ), Eqs. (\ref{eq:ecc}) yield
\begin{eqnarray}
q =  \frac{1}{2}~ a~ \left [ 1 + 2 \sin \left (\frac{2\pi t}{P}\right) \sqrt{e_{in}(1-e_{in})} - \cos\left(\frac{2\pi t}{P}\right)(2e_{in}-1)\right] ~.
\end{eqnarray}

\subsection{Distribution and density functions of perihelion distance}
We are interested in the probability $P( q < q_{L} )$ that a perihelion distance $q$ is less than a given value $q_{L}$. 
Since $q$ $=$ $a$ ( 1 $-$ $e$ ) and semi-major axis $a$ is a constant in secular evolution, we immediately
obtain $P( e < e_{L} )$ $=$ $P( q > q_{L} )$. Thus,
\begin{eqnarray}\label{eq:qp}
P( q < q_{L} ) &=& 1 ~-~ P (  e < e_{L} ) ~,
\nonumber \\
F_{q} (q) &=& 1 ~-~ F(e) ~,
\nonumber \\
P( q < q_{L} ) &=& F_{q} (q_{L}) ~,
\nonumber \\
e &=& 1 ~-~ \frac{q}{a}  ~,
\nonumber \\
q &\in& \langle 0, a \rangle ~,
\end{eqnarray}
where $F_{q}$ is the distribution function of perihelion distance.

The density function $f_{q}(q)$ of perihelion distance $q$ is, on the basis of Eqs. (\ref{eq:qp}), 
\begin{eqnarray}\label{eq:qpdf}
f_{q}( q) ~dq &=& -~ f(e) ~de ~,
\nonumber \\
f_{q} (q) &=&  \frac{1}{a} ~ f(e)  ~,
\nonumber \\
e &=& 1 ~-~ \frac{q}{a}  ~,
\nonumber \\
q &\in& \langle 0, a \rangle ~.
\end{eqnarray}

Eqs. (\ref{eq:f1})-(\ref{eq:f2}) and Eqs. (\ref{eq:qpdf}) yield for the density function $f_{q} (q)$:
\begin{eqnarray}\label{eq:fq1}
f_{q} (q) &=&  \frac{1}{a} ~ f(e)  ~,~~ q \in \langle 0, a \rangle ~,
\nonumber \\
f (e) &=& f \left [ g^{-1} \left ( e \right ) \right ]
\frac{1}{\left | \cos \chi ~+~ X_{e1} \right |} ~,
\nonumber \\
X_{e1} &=& \frac{(2e-1) \cos \chi ~ \sin \chi  ~-~ 2 \sqrt{e(1-e)} \left | \sin \chi \right | \sin \chi}{
\sqrt{1 + \left [ 1 + 8 e (e-1) \right ] \sin^2 \chi - (2e-1)^2~+~ 4 ( 2 e - 1 ) \sqrt{e(1-e)} \left|\sin \chi \right|\cos\chi}} ~,
\nonumber \\
e_{in} &=& g^{-1}(e) = -~ \sqrt{e(1-e)} \left | \sin\chi \right | + \frac{1}{2} \left [ 1 + (2e-1) \cos \chi \right ] ~,
\nonumber \\
e &=& 1 ~-~ \frac{q}{a}  ~,
\nonumber \\
\chi &=& \frac{2~\pi}{P} ~t ~,
\nonumber \\
e_{in} &\in& \langle 0, \sin^2 \left ( \frac{\pi t}{P}\right ) \rangle ~,
\end{eqnarray}
and,
\begin{eqnarray}\label{eq:fq2}
f_{q} (q) &=&  \frac{1}{a} ~ f(e)  ~, ~~ q \in \langle 0, a \rangle ~,
\nonumber \\
f(e) &=& f \left [ g^{-1} \left ( e \right ) \right ]
\frac{1}{ \left | \cos \chi ~+~ X_{e2} \right | } ~,
\nonumber \\
X_{e2} &=& \frac{(2e-1) \cos \chi ~ \sin \chi  ~+~ 2 \sqrt{e(1-e)} \sin^2 \chi}{\sqrt{1 + \left [ 1 + 8 e (e-1) \right ] \left | \sin \chi \right | \sin \chi - (2e-1)^2 ~-~ 4 ( 2 e - 1 ) \sqrt{e(1-e)} \left|\sin \chi \right|\cos\chi}} ~,
\nonumber \\
e_{in} &=& g^{-1}(e) = +~ \sqrt{e(1-e)} \left | \sin \chi \right | + \frac{1}{2} \left [ 1 + (2e-1) \cos \chi \right ] ~,
\nonumber \\
e &=& 1 ~-~ \frac{q}{a}  ~,
\nonumber \\
\chi &=& \frac{2~\pi}{P} ~t ~,
\nonumber \\
e_{in} &\in& \langle \sin^2 \left ( \frac{\pi t}{P} \right), 1 \rangle ~.
\end{eqnarray}

The probability $P( q < q_{L} )$ that a perihelion distance $q$ is less than a given value $q_{L}$ can be calculated from
\begin{eqnarray}\label{eq:qpf}
P( q < q_{L} ) &=& \int_{0}^{q_{L}} ~ f_{q} (q') ~dq'  ~,
\nonumber \\
P( q < q_{L} ) &=& F_{q} (q_{L})  ~.
\end{eqnarray}
where Eqs. (\ref{eq:fq1})-(\ref{eq:fq2}) are to be used.

\subsubsection{Uniform distribution}
Let us consider, as an example, a uniform distribution of eccentricities.
The density and distribution functions of perihelion distance are, on the basis of Eqs. (\ref{eq:ud1}), (\ref{eq:qpdf}) and Eqs. (\ref{eq:qpf}),
\begin{eqnarray}\label{eq:udq1}
f_{q} (q_{in}) &=& \frac{1}{a} ~,  
\nonumber \\
F_{q} (q_{in}) &=& \int_{0}^{q_{in}} ~f_{q} (q_{in}')~ dq_{in}' = q_{in} / a ~,
\nonumber \\
q_{in} &\in& \langle 0, a \rangle ~.
\end{eqnarray}
The distribution function $F(q_{in})$ fulfills the condition $F_{q}(a)$ $=$ 1.

Using Eqs. (\ref{eq:fq1})-(\ref{eq:fq2}), we obtain
\begin{eqnarray}\label{eq:udfq1}
f (e) &=& \frac{1}{a} ~ \frac{1}{\left | \cos \chi ~+~ X_{e1} \right |} ~,
\nonumber \\
X_{e1} &=& \frac{(2e-1) \cos \chi ~ \sin \chi  ~-~ 2 \sqrt{e(1-e)} \left | \sin \chi \right | \sin \chi}{
\sqrt{1 + \left [ 1 + 8 e (e-1) \right ] \sin^2 \chi - (2e-1)^2~+~ 4 ( 2 e - 1 ) \sqrt{e(1-e)} \left|\sin \chi \right|\cos\chi}} ~,
\nonumber \\
e &=& 1 ~-~ \frac{q}{a}  ~,
\nonumber \\
\chi &=& \frac{2~\pi}{P} ~t ~,
\nonumber \\
e_{in} &\in& \langle 0, \sin^2 \left ( \frac{\pi t}{P}\right ) \rangle ~,
\end{eqnarray}
and,
\begin{eqnarray}\label{eq:udfq2}
f(e) &=& \frac{1}{a} ~\frac{1}{ \left | \cos \chi ~+~ X_{e2} \right | } ~,
\nonumber \\
X_{e2} &=& \frac{(2e-1) \cos \chi ~ \sin \chi  ~+~ 2 \sqrt{e(1-e)} \sin^2 \chi}{\sqrt{1 + \left [ 1 + 8 e (e-1) \right ]  \left | \sin \chi \right | \sin \chi - (2e-1)^2 ~-~ 4 ( 2 e - 1 ) \sqrt{e(1-e)} \left|\sin \chi \right|\cos\chi}} ~,
\nonumber \\
e &=& 1 ~-~ \frac{q}{a}  ~,
\nonumber \\
\chi &=&  \frac{2~\pi}{P} ~t ~,
\nonumber \\
e_{in} &\in& \langle \sin^2 \left ( \frac{\pi t}{P} \right), 1 \rangle ~.
\end{eqnarray}

As an illustration we take $t$ $=$ $P /2$. Eqs. (\ref{eq:udfq1})-(\ref{eq:udfq2}) yield $f_{q}(q)$ $=$ $1/a$, and, the distribution function is $F_{q}(q)$ $=$ $q/a$:
\begin{eqnarray}\label{eq:udfq3}
f_{q}(q) &=& \frac{1}{a} ~, ~~ t = 0 ~,
\nonumber \\
F_{q}(q) &=& \frac{q}{a} ~, ~~ t = 0 ~,
\nonumber \\
f_{q}(q) &=& \frac{1}{a} ~, ~~ t = \frac{P}{2} ~,
\nonumber \\
F_{q}(q) &=& \frac{q}{a} ~, ~~ t = \frac{P}{2} ~.
\end{eqnarray}
Eqs. (\ref{eq:qpf}) may be used.

\subsubsection{Distribution function $F(e_{in})$ $=$ $e_{in}^{2}$}
We will consider, as an another example, $F(e_{in})$ $=$ $e_{in}^{2}$ (see Sec. 5.1.2). The corresponding density 
and distribution functions of perihelion distance are, on the basis of Eqs. (\ref{eq:jeans}), (\ref{eq:qpdf}) and Eqs.(\ref{eq:qpf}),
\begin{eqnarray}\label{eq:jeansq}
f_{q}(q_{in}) &=& 2~\frac{1}{a} ~ \left ( 1 ~-~ \frac{q_{in}}{a} \right ) ~, 
\nonumber \\
F_{q} (q_{in}) &=& \int_{0}^{q_{in}} ~f_{q} (q_{in}')~ dq_{in}' = \frac{q_{in}}{a}  ~ \left ( 2~-~ \frac{q_{in}}{a} \right ) ~,
\nonumber \\
q_{in} &\in& \langle 0, a \rangle ~.
\end{eqnarray}

Using Eqs. (\ref{eq:fq1})-(\ref{eq:fq2}), we obtain
\begin{eqnarray}\label{eq:jeansfq1}
f_{q} (q) &=&  \frac{1}{a} ~ f(e)  ~,~~ q \in \langle 0, a \rangle ~,
\nonumber \\
f (e) &=& \frac{1+(2e-1) \cos \chi ~-~ 2 \sqrt{e(1-e)} \left |\sin \chi \right|}{\left | \cos \chi ~+~ X_{e1} \right |} ~,
\nonumber \\
X_{e1} &=& \frac{(2e-1) \cos \chi ~ \sin \chi  ~-~ 2 \sqrt{e(1-e)} \left | \sin \chi \right | \sin \chi}{
\sqrt{1 + \left [ 1 + 8 e (e-1) \right ] \sin^2 \chi - (2e-1)^2~+~ 4 ( 2 e - 1 ) \sqrt{e(1-e)} \left|\sin \chi \right|\cos\chi}} ~,
\nonumber \\
e &=& 1 ~-~ \frac{q}{a}  ~,
\nonumber \\
\chi &=& \frac{2~\pi}{P} ~t ~,
\nonumber \\
e_{in} &\in& \langle 0, \sin^2 \left ( \frac{\pi t}{P}\right ) \rangle ~,
\end{eqnarray}
and,
\begin{eqnarray}\label{eq:jeansfq2}
f_{q} (q) &=&  \frac{1}{a} ~ f(e)  ~,~~ q \in \langle 0, a \rangle ~,
\nonumber \\
f(e) &=& \frac{1+(2e-1) \cos \chi  ~+~ 2 \sqrt{e(1-e)} \left |\sin \chi \right|}{ \left | \cos \chi ~+~ X_{e2} \right | } ~,
\nonumber \\
X_{e2} &=& \frac{(2e-1) \cos \chi ~ \sin \chi  ~+~ 2 \sqrt{e(1-e)} \sin^2 \chi}{\sqrt{1 + \left [ 1 + 8 e (e-1) \right ]  
\left | \sin \chi \right | \sin \chi - (2e-1)^2 ~-~ 4 ( 2 e - 1 ) \sqrt{e(1-e)} \left|\sin \chi \right|\cos\chi}} ~,
\nonumber \\
e &=& 1 ~-~ \frac{q}{a}  ~,
\nonumber \\
\chi &=&  \frac{2~\pi}{P} ~t ~,
\nonumber \\
e_{in} &\in& \langle \sin^2 \left ( \frac{\pi t}{P} \right), 1 \rangle ~.
\end{eqnarray}

As an illustration we take $t$ $=$ $P /2$. Eqs. (\ref{eq:jeansfq1})-(\ref{eq:jeansfq2}) yield $f_{q}(q)$ $=$ 2 $q / a^{2}$, and, 
the distribution function is $F_{q}(q)$ $=$ $(q/a)^{2}$:
\begin{eqnarray}\label{eq:jeansfq3}
f_{q}(q) &=& 2~\frac{1}{a} ~ \left ( 1 ~-~ \frac{q}{a} \right ) ~, ~~ t = 0 ~, 
\nonumber \\
F_{q} (q) &=&  \frac{q}{a}  ~ \left ( 2~-~ \frac{q}{a} \right ) ~, ~~ t = 0 ~,
\nonumber \\
f_{q}(q) &=& 2~\frac{1}{a} ~  \frac{q}{a} ~, ~~ t = \frac{P}{2} ~,
\nonumber \\
F_{q}(q) &=& \left (  \frac{q}{a} \right )^{2} ~, ~~ t = \frac{P}{2} ~,
\nonumber \\
q &\in& \langle 0, a \rangle ~.
\end{eqnarray}
The situation represented by Eqs. (\ref{eq:jeansfq3}) is illustrated in Fig. 7. 

\begin{figure}[h]
\centering
\includegraphics[scale=0.65]{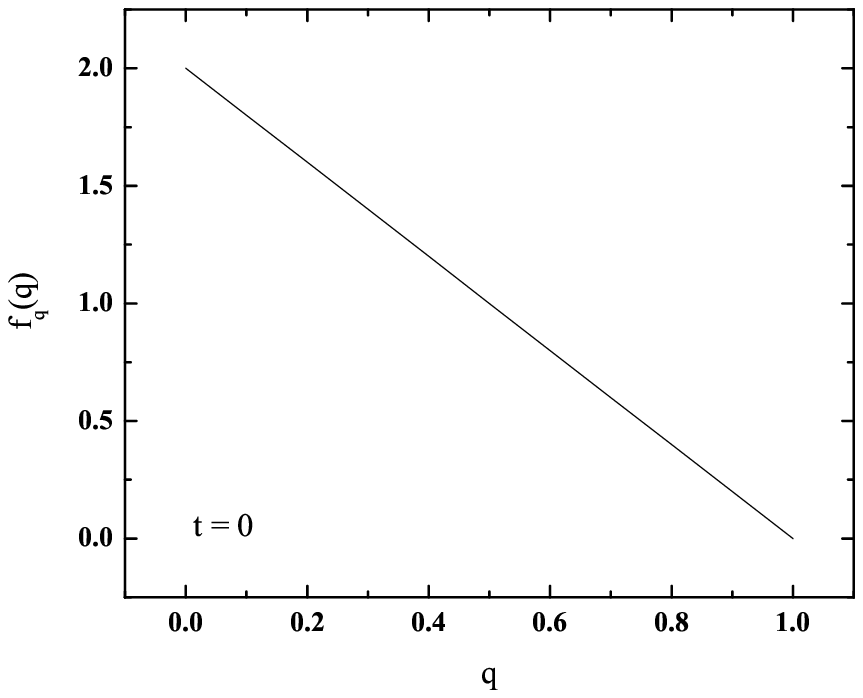}
\includegraphics[scale=0.65]{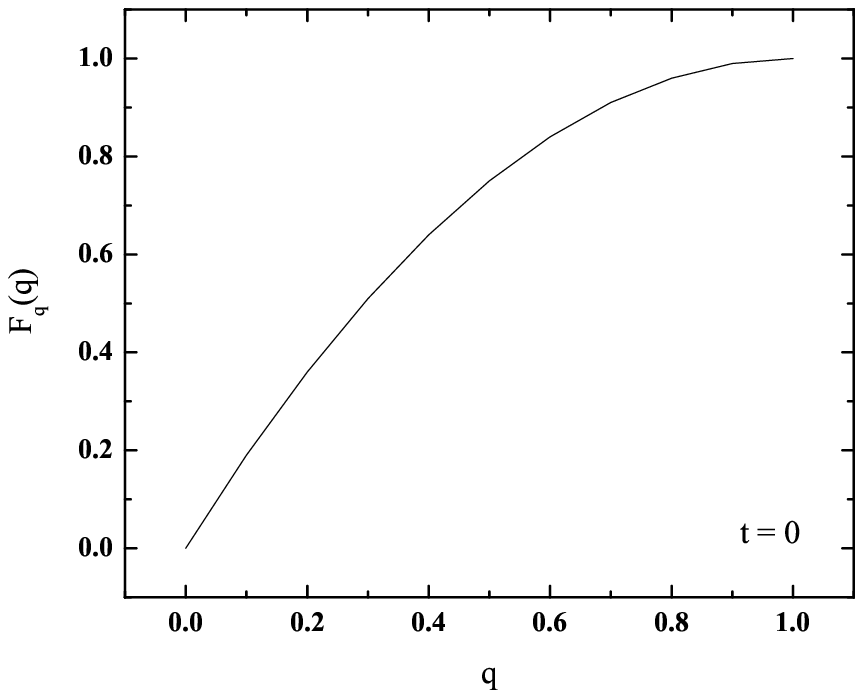}
\includegraphics[scale=0.65]{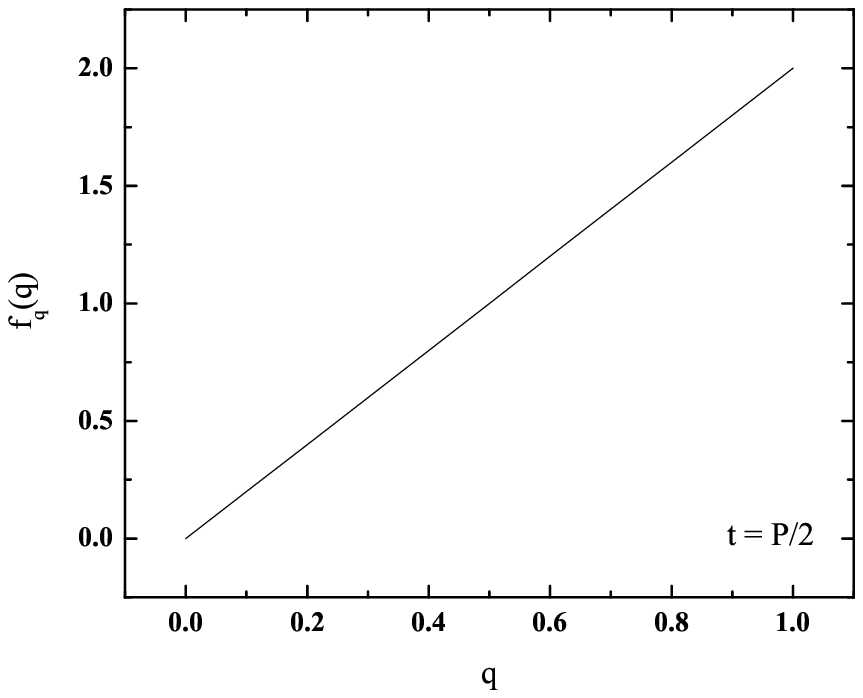}
\includegraphics[scale=0.65]{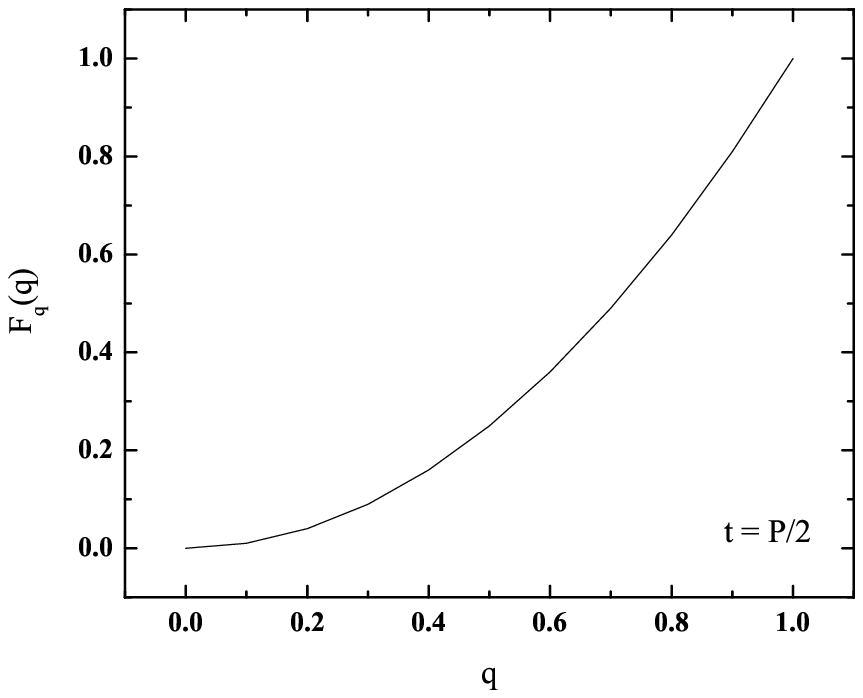}
\label{F7}
\caption{Density $f_{q}(q)$ and distribution $F_{q}(q)$ functions of perihelion distance for two instants: 
$t$ $=$ 0 and $t$ $=$ $P/2$, where $P$ is the oscillation period of eccentricities 
(perihelion and aphelion distances, angular orbital elements).
Galactic tide and gravity of the Sun play role. Semi-major axis is normed to 1, for simplicity.}
\end{figure}

\subsection{$P(q < q_{L})$ for a random time}
Orbital evolution of perihelion distance is given by the evolution of eccentricity, since
the relation between the semi-major axis $a$ and eccentricity $e$ is $q$ $=$ $a$ ( 1 $-$ $e$ ), and
$a$ is a constant in secular evolution due to the tidal effect of the Galaxy. We have, on the basis
of Eqs. (\ref{eq:ecc}),
\begin{eqnarray}\label{eq:peri1}
q ( t ) &=& \frac{a}{2} \left \{ 1 ~+~ \sin \left (  \frac{2~\pi}{P} ~t ~-~ \frac{\pi}{2}~\delta_{in}  \right ) \right \} ~,
\nonumber \\
\delta_{in} &\in& \langle  -~1, ~+ 1 \rangle ~. 
\end{eqnarray}
The probability $P(q < q_{L})$ that a value $q$ is smaller than a value $q_{L}$, in the sense of Sec. 5.2, is
\begin{equation}\label{eq:peri2}
P(q < q_{L}) = \frac{1}{2} ~-~ \frac{1}{\pi} ~\arcsin \left ( 1 ~-~ 2~\frac{q_{L}}{ a} \right ) ~.
\end{equation}
If $q_{L} / a$ $\ll$ 1, then Eq. (\ref{eq:peri2}) reduces to 
\begin{equation}\label{eq:peri3}
P(q < q_{L}) = \frac{2}{\pi} ~ \sqrt{\frac{q_{L}}{ a }}  ~, ~~ q_{L} / a \ll 1 ~.
\end{equation}
If one uses the form $q$ $=$ $a$
$\bigl |1 ~-~ 2~ | t ~-~ t_{in} | (mod~ P) / P \bigr |$
instead of Eq. (\ref{eq:peri1}), then
$P(q < q_{L})$ $=$ $q_{L} / a$.
Eqs. (\ref{eq:peri2})-(\ref{eq:peri3}) hold for the case when gravity of the Sun and galactic tide are considered.

\subsection{Distribution of perihelion distances -- galactic tides and results of numerical calculations}
Fig. 8 illustrates distribution of perihelion distances, histogram and cumulative number of comets $N( < q )$ as a function 
of a cometary perihelion distance $q$. The results presented in Fig. 8 correspond to the gravity of the Sun and the effect of galactic tide, only.
Numerical integrations of equation of motion for 4.5 $\times$ 10$^{9}$ years for semi-major axis $a_{in}$ $=$ 5 $\times$ 10$^{4}$ AU, 
inclination with respect to the galactic equatorial plane $i_{in}$ $=$ 90 degrees and
various orientations of the initially almost circular orbits (rotation angles -- longitudes of the ascending node -- 0, 45, 90, ..., 270, 315 degrees) were performed. 

The histogram for the zone $q <$ 30 (40) AU suggests that the $q-$distribution corresponds to the 
uniform distribution and it is consistent with the observational data for the group of ``new'' comets presented 
by Fern\'{a}ndez and Gallardo (1999, Fig. 1). 

On the basis of the histogram in Fig. 8 one could come to the conclusion
that the distribution function of perihelia is given by Eq. (\ref{eq:jeansq}) and the situation
corresponds to the idea of Fern\'{a}ndez and Gallardo (1999), Hills (1981) and Jeans (1919),
as it was mentioned in Sec. 6.1.2. If the idea is physically correct, one should await that
the cumulative number of comets $N( < q )$ is given by the formula represented by Eq. (\ref{eq:jeansq}).
The least-square method fit to the data yields 
\begin{eqnarray}\label{eq:peri4}
N(<q) &=& A~ q ~+~ B~ q^{2}~,  ~~ q \ll a ~,
\nonumber \\
A &=& ( 1.08 \pm 0.03 ) ~\mbox{AU}^{- 1} ~, 
\nonumber \\
B &=& - ~( 4.9 \pm 0.3 ) \times 10^{-3} ~\mbox{AU}^{- 2} ~.
\end{eqnarray}
The comets with $a$ $=$ 5 $\times$ 10$^{4}$ AU are used. Eq. (\ref{eq:jeansq}) yields that
$A$ $=$ 2 / $a$ and $B$ $=$ $-$ 1 / $a^{2}$ and their ratio is
$| B | / A$ $=$ 1 / ( 2 $a$ ) $=$ 1 $\times$ 10$^{-5}$ AU$^{-1}$. Eq. (\ref{eq:peri4}) yields
$| B | / A$ $=$ 4.54 $\times$ 10$^{-3}$ AU$^{-1}$, i.e., 454-times greater than the value corresponding
to the idea of Jeans (1919), Hills (1981) and others. 

On the basis of Sec. 6.2 we have tried also the fit of the form $N(<q)$ $=$ $C ~q^{\gamma}$.
The least-square method yields
\begin{eqnarray}\label {eq:peri5}
N(<q) &=& C_{\gamma} ~q^{\gamma} ~, ~~ q \ll a ~,
\nonumber \\
C_{\gamma} &=& ( 3.055 \pm 0.335 ) ~ \mbox{AU}^{- \gamma} ~,
\nonumber \\
\gamma &=& 0.657 \pm 0.026 ~. 
\end{eqnarray}
The value of the exponent $\gamma$ lies between the values 1/2 and 1 discussed in Sec. 6.2.
The result suggests that $\gamma$ equals 2/3. 
If we use the least-square fit of the form $N(<q)$ $=$ $C$ $q^{2/3}$, we obtain
\begin{eqnarray}\label {eq:peri6}
N(<q) &=& C ~q^{2/3} ~, ~~ q \ll a ~,
\nonumber \\
C &=& ( 2.929 \pm 0.032  ) ~ \mbox{AU}^{- 3/2} ~.
\end{eqnarray}
The error of $C$ is 1$\%$. 

\begin{figure}[h]
\centering
\includegraphics[scale=0.65]{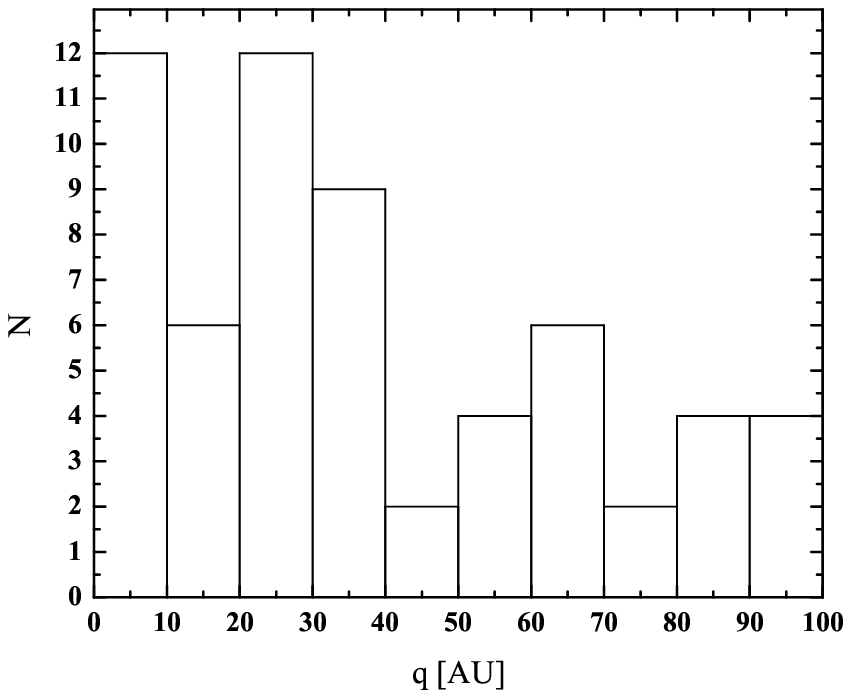}
\includegraphics[scale=0.65]{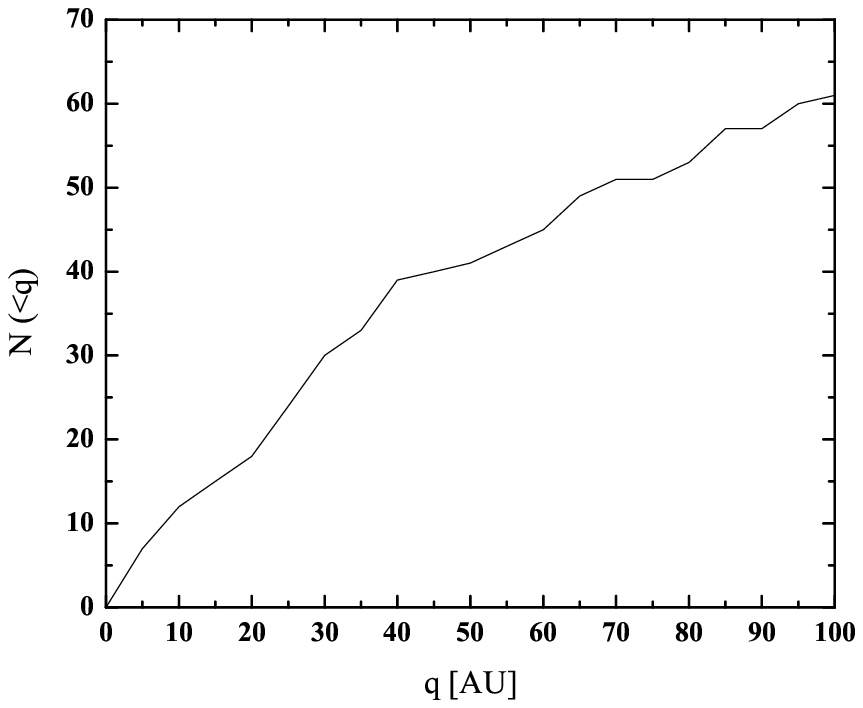}
\label{F8}
\caption{Histogram and number $N(< q)$ of comets with perihelia smaller than a value $q$ of the perihelion
distance. Only cases for the inner part of the Solar System are depicted, $q$ $<$ 100 AU.}
\end{figure}

The idea of Fern\'{a}ndez and Gallardo (1999), Hills (1981) and Jeans (1919) (and others) is based on the 
density function $f(e)$ $=$ 2 $e$ or $f_{q} (q)$ $=$ ( 2 / $a$ ) ( 1 $-$ $q / a$ ). However, also other
functions can produce uniform distribution in perihelion distances for small $q$, e.g., 
$f(e)$ $=$ ( $k$ $+$ 1) $e^{k}$, $f_{q} (q)$ $=$ [( 1 $+$ $k$ ) / $a$ ] ( 1 $-$ $q$ / $a$ )$^{k}$, 
$F_{q} (q)$ $=$ 1 $-$ ( 1 $-$ $q$ / $a$ )$^{k + 1}$ ($k$ $>$ $-$ 1); 
$F_{q} (q)$ $=$ ($k$ $+$ 1 ) $q$ / $a$ for small $q$. 
Moreover, the distribution treated by Jeans (1919) yields not only a density function for eccentricity, but also a density function for semi-major axis.
The density function for semi-major axis is, according to Jeans (1919), proportional to $\sqrt{a}$. This is not consistent
with the idea that the distribution in semi-major axis is $a^{\alpha}$, where $\alpha$ $\in$ ($-$ 4, $-$ 2) (Fern\'{a}ndez
and Gallardo 1999). Thus Eq. (\ref{eq:peri6}) may be more physical than Eq. (\ref{eq:peri4}).

\subsection{Distribution functions $F_{a,~e} (a, e)$, $F_{a,~q} (a, q)$, $F_{q} (q)$, $F_{a} (a)$}
 We want to find distribution functions $F_{a,~e} (a, e)$, $F_{a,~q} (a, q)$ and $F_{q} (q)$,
where $a$, $e$ and $q$ are the semi-major axis, eccentricity and perihelion distance.

\subsubsection{Distribution function $F_{a,~e} (a, e)$}
If the density function $f_{a,~e} (a, e)$ can be written as the product $f_{a} (a)$ $f (e)$, then
\begin{eqnarray}\label{eq:dfae1}
F_{a,~e} (a, e) &=& \int_{a_{min}}^{a} f_{a} (a ') ~ da '  ~ \int_{0}^{e} f(e ') ~de ' ~,
\nonumber \\
a &\in& \langle  a_{min}, a_{max} \rangle ~,
\nonumber \\
e &\in& \langle  0, 1 )  ~.
\end{eqnarray}
Let ($\alpha$ $\ne$ $-$ 1)
\begin{eqnarray}\label{eq:dfae2}
f_{a} (a) &=&  \left ( \alpha + 1 \right )  ~
\left (  a_{max}^{\alpha + 1} ~-~ a_{min}^{\alpha + 1} \right )^{- 1} ~a^{\alpha} ~,
\nonumber \\
f(e) &=& ( k + 1 ) ~ e^{k} ~, ~~ k > 0 ~,
\nonumber \\
a &\in& \langle  a_{min}, a_{max} \rangle ~,
\nonumber \\
e &\in& \langle  0, 1 ) ~.
\end{eqnarray}
Eqs. (\ref{eq:dfae1})-(\ref{eq:dfae2}) yield
\begin{eqnarray}\label{eq:dfae3}
F_{a,~e} (a, e) &=& \frac{ a^{\alpha + 1} ~-~ a_{min}^{\alpha + 1}}{
 a_{max}^{\alpha + 1} ~-~ a_{min}^{\alpha + 1}} ~e^{k + 1} ~,
\nonumber \\
a &\in& \langle  a_{min}, a_{max} \rangle ~,
\nonumber \\
e &\in& \langle  0, 1 ) ~.
\end{eqnarray}
The case $\alpha$ $=$ $-$ 1 would yield
\begin{eqnarray}\label{eq:dfae3-1}
f_{a} (a) &=& \left (  \ln a_{max} ~-~ \ln a_{min} \right )^{- 1} ~a^{-~1} ~,
\nonumber \\
F_{a,~e} (a, e) &=& \frac{ \ln a ~-~ \ln a_{min}}{
 \ln a_{max} ~-~ \ln a_{min}} ~e^{k + 1} ~,
\nonumber \\
a &\in& \langle  a_{min}, a_{max} \rangle ~,
\nonumber \\
e &\in& \langle  0, 1 )  ~,
\end{eqnarray}
instead of Eqs. (\ref{eq:dfae3}).

\subsubsection{Distribution function $F_{a,~q} (a, q)$}
We have to use $e$ $=$ 1 $-$ $q$ / $a$ in Sec. 6.4.1.

In general, we have
\begin{eqnarray}\label{eq:dfaq0}
F_{a,~q} (a, q) &=& \int_{a_{min}}^{a} ~ \int_{1 - q / a_{1}}^{1} f_{a,~e} (a_{1}, e) ~de ~ da_{1} ~,
\nonumber \\
a &\in& \langle  a_{min}, a_{max} \rangle ~,
\nonumber \\
q &\in& ( 0, a_{min} \rangle ~,
\nonumber \\
F_{a,~q} (a, q) &=& \int_{a_{min}}^{q} ~ \int_{0}^{1} f_{a,~e} (a_{1}, e) ~de ~ da_{1} 
\nonumber \\
&& +~ \int_{q}^{a} ~ \int_{1 - q / a_{1}}^{1} f_{a,~e} (a_{1}, e) ~de ~ da_{1} ~,
\nonumber \\
a &\in& \langle  a_{min}, a_{max} \rangle ~,
\nonumber \\
q &\in& \langle a_{min}, a \rangle ~.
\end{eqnarray}

If we consider the conventional approach $k$ $=$ 1
(e.g., Hills 1981, Fern\'{a}ndez and Gallardo 1999),
then Eqs. (\ref{eq:dfaq0}) yield
\begin{eqnarray}\label{eq:dfaq1}
F_{a,~q} (a, q) &=& \frac{\alpha + 1}{a_{max}^{\alpha + 1} ~-~ a_{min}^{\alpha + 1}} ~ \left \{
\frac{2}{\alpha} ~ q~ \left  (  a^{\alpha} ~-~ a_{min}^{\alpha} \right ) ~-~
\frac{1}{\alpha - 1}  ~ q^{2} ~ \left ( a^{\alpha - 1} ~-~ a_{min}^{\alpha - 1} \right ) \right \} ~,
\nonumber \\
a &\in& \langle  a_{min}, a_{max} \rangle ~,
\nonumber \\
q &\in& ( 0, a_{min} \rangle ~,
\nonumber \\
F_{a,~q} (a, q) &=& \frac{q^{\alpha + 1} ~-~ a_{min}^{\alpha + 1}}{a_{max}^{\alpha + 1} ~-~ a_{min}^{\alpha + 1}} 
\nonumber \\
& & +~ \frac{\alpha + 1}{a_{max}^{\alpha + 1} ~-~ a_{min}^{\alpha + 1}} ~ \left \{
\frac{2}{\alpha} ~ q~ \left  (  a^{\alpha} ~-~ q^{\alpha} \right ) ~-~
\frac{1}{\alpha - 1}  ~ q^{2} ~ \left ( a^{\alpha - 1} ~-~ q^{\alpha - 1} \right ) \right \} ~,
\nonumber \\
a &\in& \langle  a_{min}, a_{max} \rangle ~,
\nonumber \\
q &\in& \langle a_{min}, a \rangle ~.
\end{eqnarray}

If $k$ is more general, not only the special case $k$ $=$ 1, then Eqs. (\ref{eq:dfaq0}) yield  ($\alpha$ $\ne$ $-$ 1)
\begin{eqnarray}\label{eq:dfaq5}
F_{a,~q} (a, q) &=& \frac{\alpha + 1}{a_{max}^{\alpha + 1} ~-~ a_{min}^{\alpha + 1}} ~ 
\int_{a_{min}}^{a} a_{1}^{\alpha} ~
\left \{  ~1 ~-~  \left (  1~-~  \frac{q}{a_{1}} \right )^{k + 1} \right \} ~da_{1}  = 
\nonumber \\
&=& \frac{\alpha + 1}{a_{max}^{\alpha + 1} ~-~ a_{min}^{\alpha + 1}} ~
\sum_{l = 1}^{k + 1} {k+1 \choose l} ~ \frac{\left ( - ~ 1 \right )^{l + 1}}{\alpha+ 1 - l} ~q^{l} ~
\left ( a^{\alpha + 1 - l} - a_{min}^{\alpha + 1 - l} \right ) ~,
\nonumber \\
q &\in& \langle 0, a_{min} \rangle ~,
\nonumber \\
a &\in& \langle  a_{min}, a_{max} \rangle ~,
\nonumber \\
F_{a,~q} (a, q) &=& \frac{q^{\alpha + 1} ~-~ a_{min}^{\alpha + 1}}{a_{max}^{\alpha + 1} ~-~ a_{min}^{\alpha + 1}} 
\nonumber \\
& & +~ \frac{\alpha + 1}{a_{max}^{\alpha + 1} ~-~ a_{min}^{\alpha + 1}} ~ 
\int_{q}^{a} a_{1}^{\alpha} ~
\left \{  ~1 ~-~  \left (  1~-~  \frac{q}{a_{1}} \right )^{k + 1} \right \} ~da_{1}  = 
\nonumber \\
&=& \frac{q^{\alpha + 1} ~-~ a_{min}^{\alpha + 1}}{a_{max}^{\alpha + 1} ~-~ a_{min}^{\alpha + 1}} 
\nonumber \\
& & +~ \frac{\alpha + 1}{a_{max}^{\alpha + 1} ~-~ a_{min}^{\alpha + 1}} ~
\sum_{l = 1}^{k + 1} {k+1 \choose l} ~ \frac{\left ( - ~ 1 \right )^{l + 1}}{\alpha+ 1 - l} ~q^{l} ~
\left ( a^{\alpha + 1 - l} - q^{\alpha + 1 - l} \right ) ~,
\nonumber \\
q &\in& \langle a_{min}, a \rangle ~,
\nonumber \\
a &\in& \langle  a_{min}, a_{max} \rangle ~.
\end{eqnarray}
Of course, $k$ may not be an integer number.

The conventional statements on the Oort cloud are: $\alpha$ $=$ $-$ 3/2 or $\alpha$ $\in$ ( $-$ 4, $-$ 2 ) and $f(e)$ $=$ 2 $e$
(e.g., Duncan et al. 1987, Bailey 1983, Fern\'{a}ndez and Ip 1987,
Fern\'{a}ndez 1992, Fern\'{a}ndez and Gallardo 1999). We have obtained some results on the values of $\alpha$
and $a_{min}$ already in Sec. 2, see the last paragraph of Sec. 2. Also results from Secs. 9 and 10 will be helpful.

\subsubsection{Distribution function $F_{q} (q)$}
We have found that $F_{a,~q} (a, q)$ is represented by Eqs. (\ref{eq:dfaq1})
if $f(e)$ $=$ 2 $e$. If we are interested in $F_{q} (q)$, then we have to consider
$a$ $\in$ $\langle a_{min}, a_{max} \rangle$. We obtain
$F_{q} (q)$ $=$ $F_{a,~q} (a_{max}, q)$, or
\begin{eqnarray}\label{eq:dfaq7}
F_{q} (q) &=& \frac{\alpha + 1}{a_{max}^{\alpha + 1} ~-~ a_{min}^{\alpha + 1}} ~ \left \{
\frac{2}{\alpha} ~ q~ \left  (  a_{max}^{\alpha} ~-~ a_{min}^{\alpha} \right ) ~-~
\frac{1}{\alpha - 1}  ~ q^{2} ~ \left ( a_{max}^{\alpha - 1} ~-~ a_{min}^{\alpha - 1} \right ) \right \} ~,
\nonumber \\
q &\in& \langle 0, a_{min} \rangle ~,
\nonumber \\
F_{q} (q) &=& \frac{q^{\alpha + 1} ~-~ a_{min}^{\alpha + 1}}{a_{max}^{\alpha + 1} ~-~ a_{min}^{\alpha + 1}} 
\nonumber \\
& & +~ \frac{\alpha + 1}{a_{max}^{\alpha + 1} ~-~ a_{min}^{\alpha + 1}} ~ \left \{
\frac{2}{\alpha} ~ q~ \left  (  a_{max}^{\alpha} ~-~ q^{\alpha} \right ) ~-~
\frac{1}{\alpha - 1}  ~ q^{2} ~ \left ( a_{max}^{\alpha - 1} ~-~ q^{\alpha - 1} \right ) \right \} ~,
\nonumber \\
q &\in& \langle a_{min}, a_{max} \rangle ~.
\end{eqnarray}
Eq. (\ref{eq:dfaq7}) improves the result of, e.g., Hills (1981), Fern\'{a}ndez and Gallardo (1999).

The case of arbitrary $k$, given by Eqs. (\ref{eq:dfaq5}), leads to
\begin{eqnarray}\label{eq:dfaq8}
F_{q} (q) &=& \frac{\alpha + 1}{a_{max}^{\alpha + 1} ~-~ a_{min}^{\alpha + 1}} ~ 
\int_{a_{min}}^{a_{max}} a_{1}^{\alpha} ~
\left \{  ~1 ~-~  \left (  1~-~  \frac{q}{a_{1}} \right )^{k + 1} \right \} ~da_{1}  = 
\nonumber \\
&=& \frac{\alpha + 1}{a_{max}^{\alpha + 1} ~-~ a_{min}^{\alpha + 1}} ~
\sum_{l = 1}^{k + 1} {k+1 \choose l} ~ \frac{\left ( - ~ 1 \right )^{l + 1}}{\alpha+ 1 - l} ~q^{l} ~
\left ( a_{max}^{\alpha + 1 - l} - a_{min}^{\alpha + 1 - l} \right ) ~,
\nonumber \\
q &\in& \langle 0, a_{min} \rangle ~,
\nonumber \\
F_{q} (q) &=& \frac{q^{\alpha + 1} ~-~ a_{min}^{\alpha + 1}}{a_{max}^{\alpha + 1} ~-~ a_{min}^{\alpha + 1}} 
\nonumber \\
& & +~ \frac{\alpha + 1}{a_{max}^{\alpha + 1} ~-~ a_{min}^{\alpha + 1}} ~ 
\int_{q}^{a_{max}} a_{1}^{\alpha} ~
\left \{  ~1 ~-~  \left (  1~-~  \frac{q}{a_{1}} \right )^{k + 1} \right \} ~da_{1}  = 
\nonumber \\
&=& \frac{q^{\alpha + 1} ~-~ a_{min}^{\alpha + 1}}{a_{max}^{\alpha + 1} ~-~ a_{min}^{\alpha + 1}} 
\nonumber \\
& & +~ \frac{\alpha + 1}{a_{max}^{\alpha + 1} ~-~ a_{min}^{\alpha + 1}} ~
\sum_{l = 1}^{k + 1} {k+1 \choose l} ~ \frac{\left ( - ~ 1 \right )^{l + 1}}{\alpha+ 1 - l} ~q^{l} ~
\left ( a_{max}^{\alpha + 1 - l} - q^{\alpha + 1 - l} \right ) ~,
\nonumber \\
q &\in& \langle a_{min}, a_{max} \rangle ~.
\end{eqnarray}

As a consequence, $F_{q} (q)$ $\propto$ $q^{1}$,  for small $q$, holds for arbitrary $k$, not only for $k$ $=$ 1 or $k$ integer. 
Thus, the result of observations (inner part of the Solar System) yielding $F_{q} (q)$ $\propto$ $q^{1}$ cannot be used
as an argument for $f(e)$ $=$ 2 $e$. 

\subsubsection{Distribution function $F_{a} (a)$}
The distribution and density functions $F_{a} (a)$ and $f_{a} (a)$ are
\begin{eqnarray}\label{eq:dfaaa}
F_{a} (a) &=& \frac{ a^{\alpha + 1} ~-~ a_{min}^{\alpha + 1}}{
 a_{max}^{\alpha + 1} ~-~ a_{min}^{\alpha + 1}} ~,
\nonumber \\
f_{a} (a) &=& \left ( \alpha ~+~ 1 \right ) ~\frac{ a^{\alpha}}{
 a_{max}^{\alpha + 1} ~-~ a_{min}^{\alpha + 1}} ~,
\nonumber \\
a &\in& \langle  a_{min}, a_{max} \rangle ~,
\end{eqnarray}
if Eqs. (\ref{eq:dfae3}) and relation $F_{a} (a)$ $=$ $\int_{0}^{a} f_{a} (a') ~da'$ are used.

\section{Minimal perihelion distances and inclinations for comets
with``moderate'' initial inclinations}
In order to determine some basic properties of secular orbital
evolution of a comet under the action of the galactic tide we
numerically solved the system of Eqs. (13)-(17) in
P\'{a}stor et al. (2009). We consider two values of the semi-major axis,
10 000 AU and 20 000 AU, as examples. We do not consider greater values
of the semi-major axis since the analytical approach to the secular evolution
of orbital elements does not yield results equivalent to the results obtained
by detailed solution of equation of motion (P\'{a}stor et al. 2009).

Table 3 presents numerical solutions $q_{min}$ and inclinations $i_{min}$
corresponding to the minimal perihelion distances $q_{min}$.
Also oscillation periods $P$ are given in Table 3. By the term
``inclination'' we mean the inclination with respect
to the galactic equatorial plane. Table 3 also shows initial
conditions of the numerical integrations. Initial argument of perihelion
is used only from interval $\omega_{in}$ $\in$ $\langle$0, 180$^{\circ}$).
This is is sufficient since orbital evolution for $\omega_{in}$ $=$
$\omega$ $+$ 180$^{\circ}$ is identical to the evolution
for $\omega_{in}$ $=$ $\omega$ if the values of other
initial orbital elements are fixed (P\'{a}stor et al. 2009).
We found that evolutions of eccentricity, argument of perihelion
and inclination are not very sensitive to initial value of the ascending
node for the initial inclination $i_{in}$ $\apprle$ 80$^{\circ}$ (under
the assumption that the values of other initial orbital elements are fixed).
Graphical evolutions of the ascending node for various values of $\Omega_{in}$
are approximately parallel, in this case (the parallelism does not hold for
the case $i_{in}$ $\approx$ 90$^{\circ}$). This is the reason why
the initial ascending node equals to zero for all numerical integrations
in Table 3. Table 3 also shows that the minimal perihelion distance is
a decreasing function of $i_{in}$, if the values of other initial orbital
elements are fixed. In reality, comets with initial inclinations
close to 90$^{\circ}$ can get into the interior region of the Solar System
due to the galactic tide.

\begin{table}[h t b p]
\centering
\begin{tabular}{c c c c c c c c c c}
\hline
\hline
$j$ & $a_{in}$ & $e_{in}$ & $\omega_{in}$ & $\Omega_{in}$ & $i_{in}$ &
$t_{int}$ & $q_{min}$ & $i_{min}$ & $P$\\

[-] & [AU] & [-] & [$^\circ$] & [$^\circ$] & [$^\circ$] & [10$^{10}$ years] &
[AU] & [$^\circ$] & [10$^{9}$ years]\\
\hline
1 & 10 000 & 0.4 & 0 & 0 & 30 & 2 & 4449 & 18.04 & 4.3\\
2 & 10 000 & 0.4 & 22.5 & 0 & 30 & 2 & 4565 & 19.50 & 4.6\\
3 & 10 000 & 0.4 & 45 & 0 & 30 & 2 & 4916 & 23.10 & 5.8\\
4 & 10 000 & 0.4 & 67.5 & 0 & 30 & 2 & 5496 & 27.35 & 9.7\\
5 & 10 000 & 0.4 & 90 & 0 & 30 & 2 & 5998 & 30.02 & 7.0\\
6 & 10 000 & 0.4 & 112.5 & 0 & 30 & 2 & 5497 & 27.36 & 9.7\\
7 & 10 000 & 0.4 & 135 & 0 & 30 & 2 & 4915 & 23.09 & 5.8\\
8 & 10 000 & 0.4 & 157.5 & 0 & 30 & 2 & 4565 & 19.51 & 4.6\\
9 & 10 000 & 0.4 & 0 & 0 & 60 & 2 & 1315 & 23.98 & 3.9\\
10 & 10 000 & 0.4 & 22.5 & 0 & 60 & 2 & 1351 & 25.41 & 4.7\\
11 & 10 000 & 0.4 & 45 & 0 & 60 & 2 & 1465 & 29.44 & 4.0\\
12 & 10 000 & 0.4 & 67.5 & 0 & 60 & 2 & 1614 & 33.39 & 3.1\\
13 & 10 000 & 0.4 & 90 & 0 & 60 & 2 & 1691 & 35.10 & 2.9\\
14 & 10 000 & 0.4 & 112.5 & 0 & 60 & 2 & 1613 & 33.41 & 3.1\\
15 & 10 000 & 0.4 & 135 & 0 & 60 & 2 & 1466 & 29.46 & 4.0\\
16 & 10 000 & 0.4 & 157.5 & 0 & 60 & 2 & 1348 & 25.37 & 4.7\\
17 & 20 000 & 0.4 & 0 & 0 & 30 & 2 & 8722 & 17.99 & 1.5\\
18 & 20 000 & 0.4 & 22.5 & 0 & 30 & 2 & 8973 & 19.40 & 1.6\\
19 & 20 000 & 0.4 & 45 & 0 & 30 & 2 & 9700 & 23.03 & 2.0\\
20 & 20 000 & 0.4 & 67.5 & 0 & 30 & 2 & 10915 & 27.30 & 3.4\\
21 & 20 000 & 0.4 & 90 & 0 & 30 & 2 & 11990 & 30.05 & 2.5\\
22 & 20 000 & 0.4 & 112.5 & 0 & 30 & 2 & 10921 & 27.28 & 3.4\\
23 & 20 000 & 0.4 & 135 & 0 & 30 & 2 & 9698 & 23.01 & 2.0\\
24 & 20 000 & 0.4 & 157.5 & 0 & 30 & 2 & 8968 & 19.40 & 1.6\\
25 & 20 000 & 0.4 & 0 & 0 & 60 & 2 & 2485 & 23.61 & 1.4\\
26 & 20 000 & 0.4 & 22.5 & 0 & 60 & 2 & 2572 & 25.32 & 1.6\\
27 & 20 000 & 0.4 & 45 & 0 & 60 & 2 & 2805 & 29.34 & 1.4\\
28 & 20 000 & 0.4 & 67.5 & 0 & 60 & 2 & 3116 & 33.36 & 1.1\\
29 & 20 000 & 0.4 & 90 & 0 & 60 & 2 & 3268 & 35.04 & 1.0\\
30 & 20 000 & 0.4 & 112.5 & 0 & 60 & 2 & 3109 & 33.31 & 1.1\\
31 & 20 000 & 0.4 & 135 & 0 & 60 & 2 & 2804 & 29.35 & 1.4\\
32 & 20 000 & 0.4 & 157.5 & 0 & 60 & 2 & 2573 & 25.33 & 1.6\\
\hline
\end{tabular}
\caption{Inclinations $i_{min}$ corresponding to minimal perihelion
distances  $q_{min}$ and oscillation periods $P$ as a result
of numerical integration of the system of differential
equations for secular evolution of orbital elements. Time
of integration is $t_{int}$ 2 $\times$ 10$^{10}$ years. The initial
semi-major axis $a_{in}$, eccentricity $e_{in}$, argument of perihelion
$\omega_{in}$, longitude of the ascending node $\Omega_{in}$ and
inclination $i_{in}$ are shown.}
\label{tab:3}
\end{table}

\begin{figure}[t]
\centering
\includegraphics[scale=0.65]{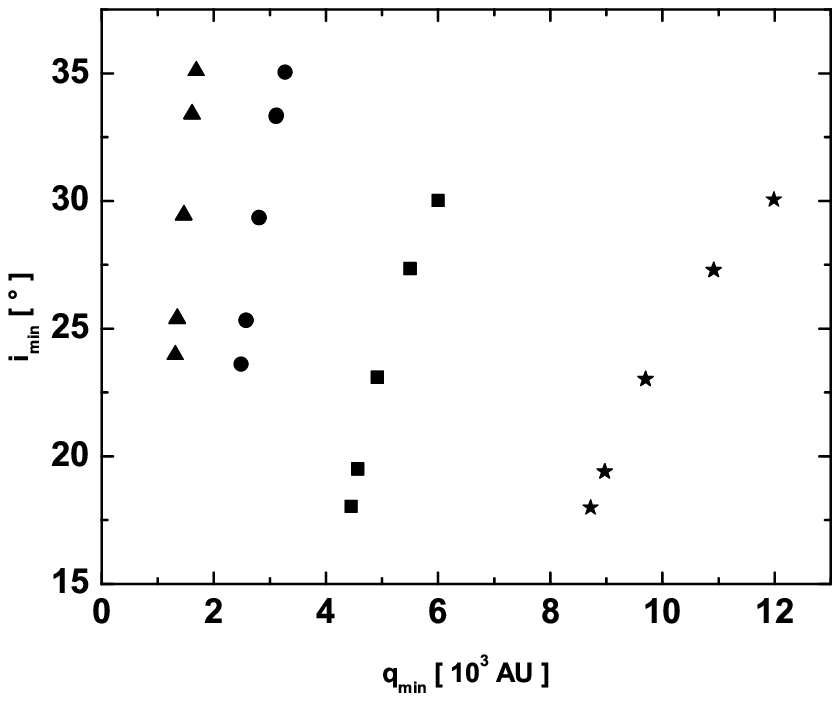}
\label{F9}
\caption{Inclination to galactic equatorial plane at minimal perihelion
distance $q_{min}$. Square is used for $a_{in}$ $=$ 10 000 AU and
$i_{in}$ $=$ 30$^{\circ}$, triangle for
$a_{in}$ $=$ 10 000 AU and $i_{in}$ $=$ 60$^{\circ}$, star for
$a_{in}$ $=$ 20 000 AU and $i_{in}$ $=$ 30$^{\circ}$ and circle for
$a_{in}$ $=$ 20 000 AU and $i_{in}$ $=$ 60$^{\circ}$. The greater $i_{in}$,
the smaller $q_{min}$ and the greater $i_{min}$, for a given $a_{in}$.}
\end{figure}

Fig. 9 depicts inclination $i_{min}$ at minimal perihelion distance
as a function of minimal perihelion distance $q_{min}$. The values are
taken from Table 3. $i_{min}$ is an increasing function of $q_{min}$
for a given $a_{in}$ and $i_{in}$. Comparison of the values of $q_{min}$ at
$a_{in}$ $=$ 10 000 AU and 20 000 AU, for a fixed $\omega_{in}$ and $i_{in}$,
yields
\begin{equation}\label{eq:qmin}
\frac{q_{min ~1}}{q_{min ~2}} \approx \frac{a_{in ~1}}{a_{in ~2}} ~.
\end{equation}
If $a_{in}$ and $i_{in}$ are fixed, then the values
$\omega_{in}$ $=$ $\omega$ and $\omega_{in}$ $=$ 180$^{\circ}$ $-$ $\omega$
yield approximately equal values of $q_{min}$ and $i_{min}$ (see also Table 3).
This is the reason why we can see only five distinct points, not eight,
in Fig. 9. This property is not related only to the values of $q_{min}$
and $i_{min}$: the complete orbital evolutions of the comets are very similar
for $\omega_{in}$ $=$ $\omega$ and $\omega_{in}$ $=$ 180$^{\circ}$ $-$
$\omega$. This property divides the interval $\omega_{in}$ $\in$
$\langle$0, 180$^{\circ}\rangle$ into two parts with different behavior.
If we consider only the values $\omega_{in}$ $\in$
$\langle$0, 90$^{\circ}\rangle$, then $q_{min}$ and
$i_{min}$ are increasing functions of $\omega_{in}$, if $a_{in}$
and $i_{in}$ are fixed. If $\omega_{in}$ $\in$ $\langle$90$^{\circ}$,
180$^{\circ}\rangle$, then $q_{min}$ and $i_{min}$ are decreasing
functions of $\omega_{in}$. An explanation of this property is probably
in the antisymmetry of the secular time derivatives after transformations
$\omega$ $\rightarrow$ $\pi$ $-$ $\omega$, $\tilde \Omega$ $\rightarrow$
$\pi$ $-$ $\tilde \Omega$ (P\'{a}stor et al. 2009). Since the orbital
evolution is not very sensitive to the initial value of the ascending node,
we obtain similar orbital evolution for $\omega_{in}$ $=$ 180$^{\circ}$
$-$ $\omega$ as for $\omega_{in}$ $=$ $\omega$.

\section{Distribution in the ecliptical inclination}
If we want to consider cometary orbital elements for the inner part of the Solar System,
then we have to consider initial inclinations with respect to the galactic equatorial
plane close to 90 degrees (see also Sec. 7). Gravity of the Sun and Galaxy will be considered.

We used numerical solution of the system of equations for secular
evolution of orbital elements given by Eqs. (13)-(17) in
P\'{a}stor et al. (2009). We used the equations to measure
minimal perihelion distance $q_{min}$ of a comet at the first three
returns of the comet to the inner part of the Solar System.
At the time when the minimal perihelion distance occurred
we measured also the inclination $i_{min}$ and the longitude
of the ascending node $\Omega_{min}$. The results are shown in Table 4.
Initial conditions of numerical integrations are also shown.
Numerical integrations with equal values of initial semi-major axes,
eccentricities and arguments of perihelion have equal oscillation period.
The oscillation period for all numerical integrations in Table 4 is
$P$ $\approx$ 1.3 $\times$ 10$^{9}$ years. Change of the longitude
of the ascending node $\Omega$ of the comet with initial inclination
$i_{in}$ $\approx$ 90$^{\circ}$ during the return of the comet
to the inner part of the Solar System is always $\approx$ $\pm \pi$.
The last column shows two ''stable'' values fulfilling the fact
that $\Omega$ lies between the values at perihelion. If the first
value is smaller/greater than the second one, then $\Omega$ increases/decreases 
from the first value to the second one. 

We are interested in an inclination $i$ of the cometary orbital plane with
respect to a reference plane. The reference plane is defined by the inclination $i_{0}$
and the longitude of the ascending node $\Omega_{0}$ with respect to the galactic 
equatorial plane (and a given reference direction). If the galactic inclination of the comet 
at its perihelion is $ i_{min}$ (galactic longitude of the ascending node $\Omega_{min}$), 
then the inclination $i$ at perihelion position can be calculated from the following equation:
\begin{equation}\label{eq:inc1}
\cos i = \cos i_{0} ~\cos i_{min} + \sin i_{0} ~\sin i_{min} ~
\cos (\Omega_{0}-\Omega_{min}) ~.
\end{equation}
Table 4 shows that the values of inclination with respect
to the galactic equator (galactic inclination) are always close
to 25$^{\circ}$ or 155$^{\circ}$. The number
of comets with galactic inclination close to 25$^{\circ}$ is approximately
equal to the number of comets with the inclination close to 155$^{\circ}$.
The values of $\Omega_{min}$ are practically random (see also Table 4).
These facts yield for a large sample of comets 
\begin{eqnarray}\label{eq:inc2}
h_{H} (i) &=& \frac{1}{2} ~ \sin i ~,
\nonumber \\
H (i) &=& \frac{1}{2} ~ \left ( 1 ~-~ \cos i \right )  ~,
\nonumber \\
\langle \cos i \rangle &=& 0 ~,
\nonumber \\
 \langle  i  \rangle &=& \frac{\pi}{2} ~.
\end{eqnarray}
The results state that the density and distribution functions correspond to isotropic distribution, the
average value of $\cos i$ equals zero and the average value of $i$ is $\pi$ / 2.
The results are consistent with Eqs. (\ref{eq:6}).

The results presented in Eqs. (\ref{eq:inc2}) hold for inclinations with respect to any reference plane
$i_{0}$ $\ne$ 0, also for the ecliptic plane. The distribution of galactic inclinations for comets
in the inner part of the Solar System is not isotropic, if only gravity of the Sun and Galaxy are
considered: $i(galactic)$ $\approx$ 25$^{\circ}$ or 155$^{\circ}$.  

\begin{table}[h t b p]
\centering
\begin{tabular}{c c c c c c c c c c c}
\hline
\hline
$j$ & $a_{in}$ & $e_{in}$ & $\omega_{in}$ & $\Omega_{in}$ & $i_{in}$ &
$q_{min}$ & $i_{min}$ & $\Omega_{min}$ & interval\\

[-] & [AU] & [-] & [$^\circ$] & [$^\circ$] & [$^\circ$] &
[AU] & [$^\circ$] & [$^\circ$] & [$^\circ$]\\
\hline
1 & 20 000 & 0.4 & 0 & 0 & 90 & 1.38 & 151.31 & 124.04 & (0,180)\\
2 & 20 000 & 0.4 & 0 & 0 & 90 & 0.82 & 24.56 & 97.06 & (180,0)\\
3 & 20 000 & 0.4 & 0 & 0 & 90 & 2.07 & 151.69 & 59.15 & (0,180)\\
4 & 20 000 & 0.4 & 0 & 45 & 90 & 2.17 & 24.80 & -34.21 & (45,-135)\\
5 & 20 000 & 0.4 & 0 & 45 & 90 & 0.39 & 153.57 & -68.24 & (-135,45)\\
6 & 20 000 & 0.4 & 0 & 45 & 90 & 0.11 & 155.36 & 128.71 & (45,225)\\
7 & 20 000 & 0.4 & 0 & 90 & 90 & 1.96 & 25.70 & -19.22 & (90,-90)\\
8 & 20 000 & 0.4 & 0 & 90 & 90 & 0.66 & 155.40 & 5.56 & (-90,90)\\
9 & 20 000 & 0.4 & 0 & 90 & 90 & 3.12 & 25.16 & 14.47 & (90,-90)\\
10 & 20 000 & 0.4 & 0 & 135 & 90 & 0.51 & 152.69 & 196.98 & (135,315)\\
11 & 20 000 & 0.4 & 0 & 135 & 90 & 0.03 & 155.43 & 411.06 & (315,495)\\
12 & 20 000 & 0.4 & 0 & 135 & 90 & 1.99 & 25.25 & 389.91 & (495,315)\\
13 & 20 000 & 0.4 & 0 & 180 & 90 & 1.20 & 153.99 & 291.56 & (180,360)\\
14 & 20 000 & 0.4 & 0 & 180 & 90 & 0.81 & 24.54 & 264.51 & (360,180)\\
15 & 20 000 & 0.4 & 0 & 180 & 90 & 1.72 & 150.34 & 234.41 & (180,360)\\
16 & 20 000 & 0.4 & 0 & 225 & 90 & 2.29 & 24.44 & 132.57 & (225,45)\\
17 & 20 000 & 0.4 & 0 & 225 & 90 & 0.45 & 154.20 & 115.59 & (45,225)\\
18 & 20 000 & 0.4 & 0 & 225 & 90 & 0.31 & 153.50 & 339.85 & (225,405)\\
19 & 20 000 & 0.4 & 0 & 270 & 90 & 1.67 & 27.66 & 150.12 & (270,90)\\
20 & 20 000 & 0.4 & 0 & 270 & 90 & 0.90 & 154.07 & 200.37 & (90,270)\\
21 & 20 000 & 0.4 & 0 & 270 & 90 & 2.79 & 26.85 & 206.01 & (290,90)\\
22 & 20 000 & 0.4 & 0 & 315 & 90 & 0.87 & 149.92 & 366.91 & (315,495)\\
23 & 20 000 & 0.4 & 0 & 315 & 90 & 0.11 & 154.00 & 606.52 & (495,675)\\
24 & 20 000 & 0.4 & 0 & 315 & 90 & 2.05 & 25.70 & 566.40 & (675,495)\\
\hline
\end{tabular}
\caption{Minimal perihelion distances $q_{min}$ during the first
three returns of a comet to the inner part of the Solar System.
Inclinations $i_{min}$ and longitudes of the ascending nodes
$\Omega_{min}$ corresponding to minimal perihelion distances
are given. Initial semi-major axis $a_{in}$, eccentricity $e_{in}$,
argument of perihelion $\omega_{in}$, longitude of the ascending node
$\Omega_{in}$ and inclination $i_{in}$ are shown. 
The longitude of the ascending node rapidly changes around $\Omega_{in}$ within
the interval presented in the last column.}
\label{T4}
\end{table}

\section{Gravity of the Galaxy, Sun and Jupiter}
This section presents results obtained when also gravity of the planet Jupiter is included.

\subsection{Evolution of orbital elements}
Evolution of semi-major axis and eccentricity is depicted in Fig. 10.
The action of the Sun and Galaxy produces constant semi-major axis of a comet in the Oort cloud, as for secular orbital evolution. 
The presence of Jupiter may cause sudden changes in semi-major axis. Moreover, the planet caused that the period of oscillations 
in eccentricity (perihelion and aphelion distances, inclination, ...) decreased to almost one half of the period found without the
action of the planet.

\begin{figure}[h]
\centering
\includegraphics[scale=0.65]{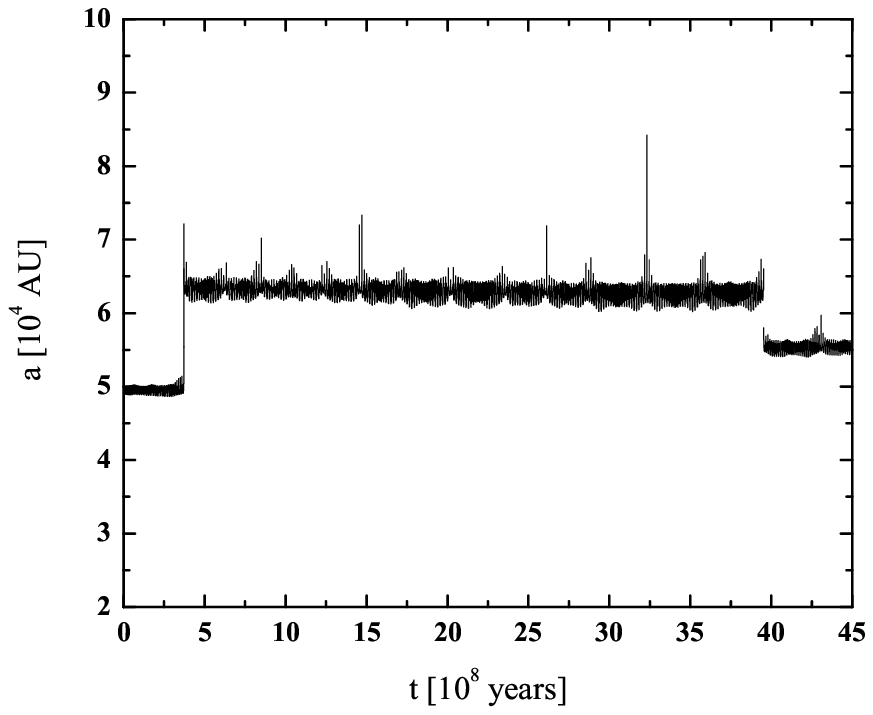}
\includegraphics[scale=0.65]{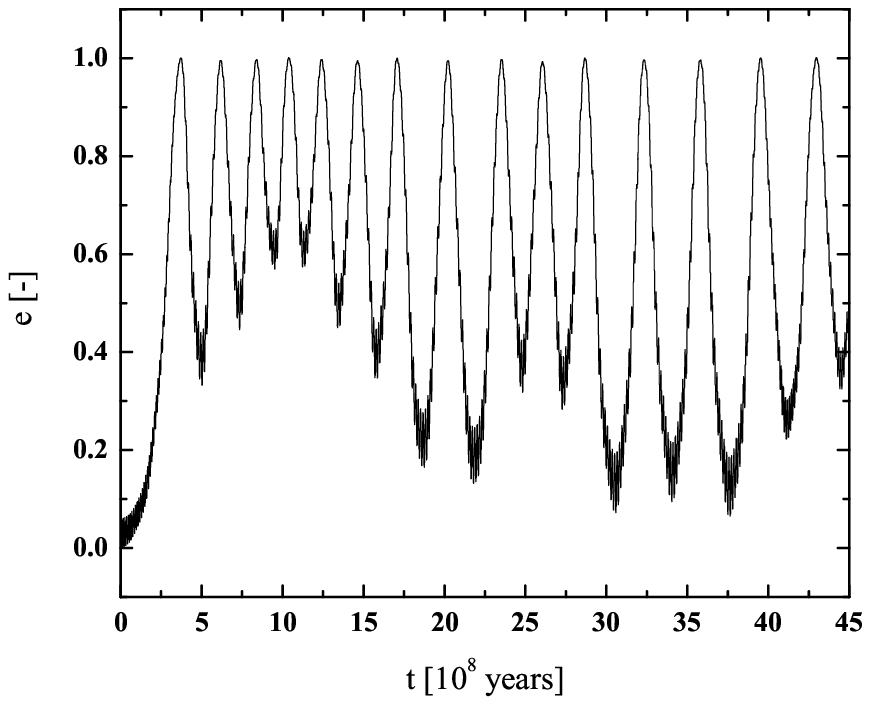}
\label{F10}
\caption{Evolution of semi-major axis and eccentricity for a comet under the gravitational influence of the Sun, Galaxy and Jupiter.}
\end{figure}

\subsection{Distribution in perihelion distance}
As it is presented in the previous subsection (see Fig. 10), the presence of Jupiter may cause more frequent returns of a comet 
to the inner part of the Solar System than it is in the case when the planet is ignored.
Moreover, our calculations confirmed the importance of the planet also in another type of computational experiment. 
We considered the same initial conditions (orbital elements -- semi-major axis 5 $\times$ 10$^{4}$ AU, 
inclination with respect to the galactic equatorial plane 90 degrees, eccentricity close to zero, ...) 
for the two cases, one without Jupiter and one with the action of Jupiter. We gathered long-period comets with perihelion distance
$q$ less than 100 AU (two groups of comets do not belong to the set of comets: i) comets with $q$ $<$ 0.01 AU, and
ii) comets ejected from the Solar System due to the close approach to Jupiter). 
While the ignorance of Jupiter yielded that 5.0$\%$ of the comets exhibited
$q$ $<$ 5 AU, the action of Jupiter, Sun and Galaxy yielded that 28.6$\%$ of the comets exhibited
$q$ $<$ 5 AU. The corresponding percentages for $q$ $<$ 10 AU are: 10.0$\%$ of the set without Jupiter
and 33.3$\%$ of the set with Jupiter. Thus, our model of the Galaxy relevantly changes
the conventional result that `planetary perturbations do not significantly alter the perihelion distance'
(see, e.g., Dones et al. 2004, p. 161).

We have already mentioned that the above presented results hold for comets which still belong to long-period comets.
Although nonnegligible part of comets was ejected from the Solar System due to the close encounter with Jupiter,
the conventional statement that 'only about 5$\%$ of the new comets are returned to Oort cloud distances
of 10$^{4}$ $-$ 10$^{5}$ AU' (Weissman 1979; Dones et al. 2004, p. 157) is not consistent with our 
numerical calculations. Our results show that most of the new comets are returned to the distances
of (10$^{4}$ $-$ 10$^{5}$) AU.

If we use the least-square fit of the form $N(<q)$ $=$ $C$ $q^{2/3}$, where $q$ is the perihelion distance
and $N(<q)$ is the number of comets with perihelion distances smaller than the value $q$, we obtain
\begin{eqnarray}\label{eq:peri6-j}
N(<q) &=& C ~q^{2/3} ~, ~~ q < 100 ~ \mbox{AU} ~,
\nonumber \\
%C &=& ( 2.2481 \pm 0.1027  ) ~ \mbox{AU}^{- 3/2} ~.
C &=& ( 2.248 \pm 0.103  ) ~ \mbox{AU}^{- 3/2} ~.
\end{eqnarray}
The error of $C$ is 4.6$\%$, greater than the error obtained without the action of Jupiter. 

\begin{figure}[h]
\centering
\includegraphics[scale=0.65]{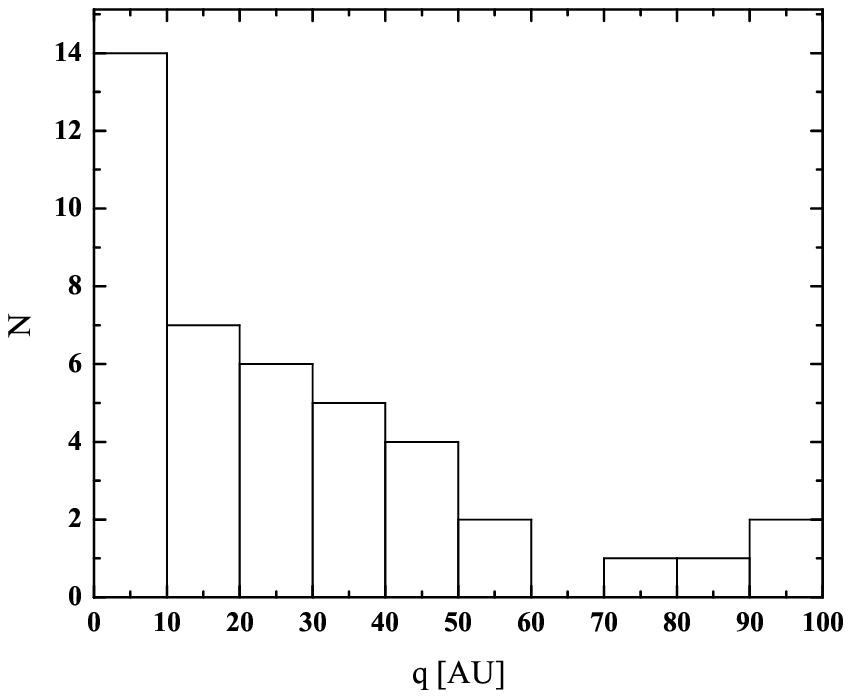}
\includegraphics[scale=0.65]{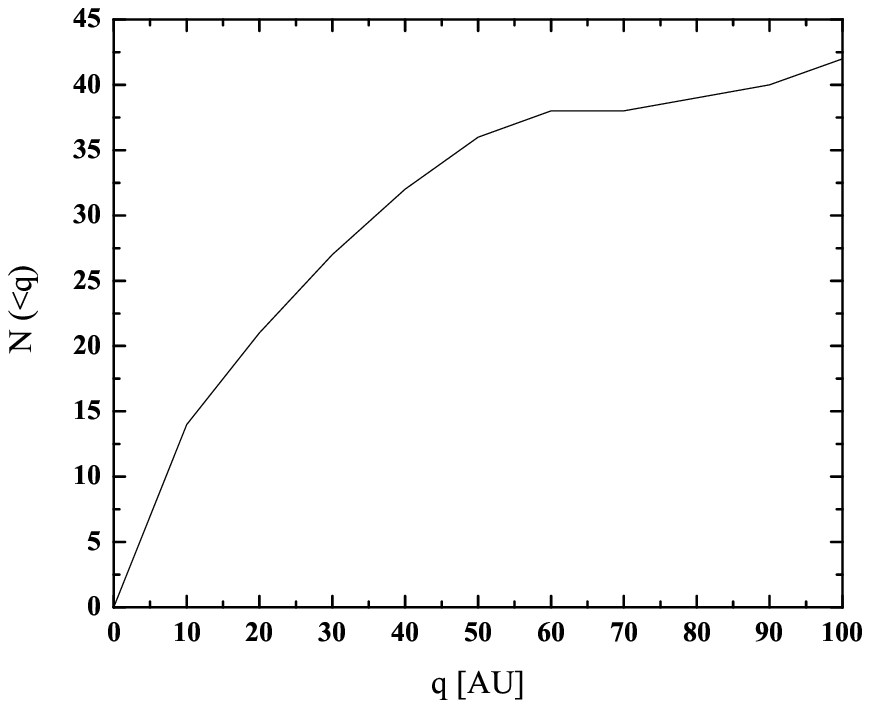}
\label{F11}
\caption{Histogram and number $N(< q)$ of comets with perihelion distances smaller than the value $q$.
Only cases for the inner part of the Solar System are depicted, $q$ $<$ 100 AU.}
\end{figure}

\subsection{Distribution in ecliptical inclination}
Fig. 12 depicts distribution in inclination with respect to the ecliptic. The action of Jupiter is
marginal, although some difference between Figs. 6 and 8 exist.
The isotropic distribution function (see Eqs. \ref{eq:5}) of ecliptical inclination -- inclination to the ecliptic --
is in a good coincidence with the calculated data. However, 70$\%$ of the calculated orbits exhibit
prograde orbits (ecliptical inclination is less than 90 degrees).

\begin{figure}[h]
\centering
\includegraphics[scale=0.65]{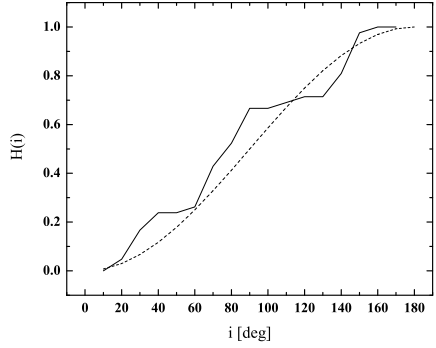}
\label{F12}
\caption{Distribution function of cometary ecliptical inclination when comets are situated at their perihelia. 
Only cases for the inner part of the Solar System are depicted, $q$ $<$ 100 AU.
The dotted line corresponds to the isotropic distribution.}
\end{figure}

One would await that gravity of a planet can influence the ecliptical inclination mainly in the cases when smaller
perihelion distances occur. If the effect of the planet would not exist, then we should await no correlation between 
the perihelion distance and the ecliptical inclination. 
Our calculations show that the inclusion of Jupiter leads to the coefficient of correlation
\begin{eqnarray}\label {eq:cor-j}
r ( q,  i_{ecl} ) &=& 0.366 ~ \pm ~ 0.135~.
\end{eqnarray}
The percentage probability that the correlation coefficient for the used set
of data is greater than the given value is less than 2$\%$.

\section{Mass of the Oort cloud}
Current estimates of the mass of the Oort cloud of comets are about (3.3 $-$ 7.0) masses of the Earth ($M_{E}$),
although the value of 38 $M_{E}$ has also appeared (Dones et al. 2004, p. 162).

\subsection{Observational data and the distribution function $F_{a} (a)$}
Using observational data on the original semi-major axes of the long-period comets taken from Marsden and William's (1997) catalogue, 
we have found that the theoretical fit for the density function $f_{a} (a)$ in the exponential form $a^{\alpha}$ holds only for
semi-major axes smaller than about 4 $\times$ 10$^{4}$ AU (see Fig. 13). Theoretical fit for the data, depicted in Fig. 13, yields
for the distribution function
\begin{eqnarray}\label {eq:nca1-j}
F_{a} (a) &=& \left (  \frac{a}{a_{0}} \right  ) ^{\alpha_{obs} + 1} ~, ~~ a \le a_{0} ~,
\nonumber \\
\alpha_{obs} ~+~ 1 &=& 1.133 \pm 0.044 ~ \mbox{AU} ~,
\nonumber \\
a_{0} &=& 4 \times 10^{4} ~ \mbox{AU} ~.
\end{eqnarray}
Relative error of the exponent is 3.9$\%$.
Comparison with Eqs. (\ref{eq:dfaaa}) yields $a_{max}$ $=$ $a_{0}$, $a_{min}$ $\longrightarrow$ 0.
Eqs. (\ref{eq:nca1-j}) give that the observed density function of semi-major axis is
\begin{eqnarray}\label {eq:nca2-j}
f_{a} (a) &=& \frac{\alpha_{obs} + 1}{a_{0}} ~\left (  \frac{a}{a_{0}} \right  ) ^{\alpha_{obs}} ~, ~~ a \le a_{0} ~,
\nonumber \\
\alpha_{obs} &=& \frac{4}{30}  ~,
\nonumber \\
a_{0} &=& 4 \times 10^{4} ~ \mbox{AU} ~.
\end{eqnarray}

\begin{figure}[h]
\centering
\includegraphics[scale=0.65]{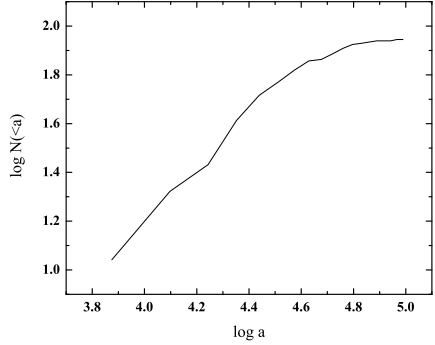}
\label{F13}
\caption{Dependence of the cumulative number of long-period comets on semi-major axis $a$ [$AU$] in logarithmic scale.
The dependence is linear for $a$ $<$ 4 $\times$ 10$^{4}$ AU.}
\end{figure}

If we would like to fit the curve in Fig. 13 up to $a$ $=$ 1 $\times$ 10$^{5}$ AU using the approximation
given by Eqs. (\ref{eq:dfaaa}), then
\begin{eqnarray}\label{eq:nca2-jn}
\alpha + 1 &=& 0.449 \pm 0.049 ~,
\nonumber \\
a_{min} &\rightarrow& 0 ~.
\end{eqnarray}

We could fit the curve in Fig. 13 up to $a$ $=$ 1 $\times$ 10$^{5}$ AU using the approximation
given by Eqs. (\ref{eq:dfaaa}) with greater values of $a_{min}$. However, large errors of the exponent $\alpha$ $+$ 1
exist in these cases. For example, the value $a_{min}$ $=$ 0.1 $\times$ 10$^{5}$ AU 
yields relative error of $\alpha$ $+$ 1: about 160$\%$.

Let us calculate mean value of the semi-major axis. On the basis of Eq. (\ref{eq:dfaaa}) we can write 
\begin{eqnarray}\label{eq:dfaaad}
\langle a \rangle  &=& \int_{a_{min}}^{a_{max}} a' ~ f_{a} (a')~da' 
\nonumber \\
&=& a_{max} ~ \frac{\alpha + 1}{\alpha + 2} ~ \frac{ 1 ~-~  \left ( a_{min} / a_{max} \right )^{\alpha + 2}}{
1 ~-~  \left ( a_{min} / a_{max} \right )^{\alpha + 1}} ~.
\end{eqnarray}

Figure 14 depicts the value of $\langle a \rangle$ / $a_{max}$ as a function of the exponent $\alpha$ for the cases
found by various authors: $\alpha$ $=$ $-$ 3/2 (Duncan et al. 1987), $\alpha$ $\in$ ( $-$ 4, $-$ 2 ) 
(Fern\'{a}ndez and Ip 1987, Fern\'{a}ndez 1992, Fern\'{a}ndez and Gallardo 1999). Also the value
$\alpha$ $=$ $-$ 0.55 given by Eqs. (\ref{eq:nca2-jn}) is considered. The value of $\langle a \rangle$ / $a_{max}$ 
for the cases $\alpha$ $=$ $-$ 3/2 and $\alpha$ $\in$ ( $-$ 4, $-$ 2 ) are calculated under the assumption that
$a_{min} / a_{max}$ $=$ 0.1 (see the solid curve and the square in Fig. 14). The case $\alpha$ $=$ $-$ 0.55
considers two possibilities, $a_{min} / a_{max}$ $=$ 0.1 (triangle in Fig. 14) and $a_{min}$ $\rightarrow$ 0
(star in Fig. 14). Our result represented by Eqs. (\ref{eq:nca2-jn}) yields the value of $\langle a \rangle$ / $a_{max}$
consistent with the value obtained from the model by Duncan et al. (1987).
The case $\alpha$ $=$ $+$ 1/2 (Jeans 1919) yields more than two times greater value than the case $\alpha$ $=$ $-$ 3/2.

\begin{figure}
\centering
 \includegraphics[scale=0.65]{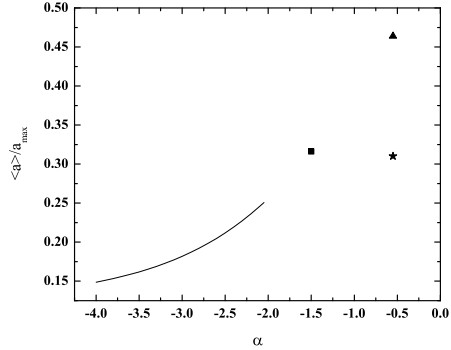}
 \caption{The ratio of the mean value of semi-major axis to $a_{max}$
 as a function of the parameter $\alpha$ for various models. If $a_{min}$ / $a_{max}$ $=$ 0.1, then the solid line 
 corresponds to the model discussed by Fern\'{a}ndez and Gallardo (1999), the square to Duncan et al. (1987), 
 the triangle to $\alpha$ $=$ $-$ 0.55. The star corresponds to Eqs. (\ref{eq:nca2-jn}) $\alpha$ $=$ $-$ 0.55 and $a_{min}$
 $\rightarrow$ 0.}
 \label{F14}
 \end{figure}

\subsection{Number of comets in the Oort cloud}
We will present two different accesses to the estimation of the number of comets in the Oort cloud.
The first one is based on the comparison between the frequencies of cometary returns followed from the new
physical model and the standard model of galactic tides. The second access considers the distribution of comets
in the semi-major axis $a$ when the density function is taken in the form proportional to $a^{\alpha}$. 

\subsubsection{The first access}
We have already mentioned the relevance of the physical model of the galactic tide 
(see Sec. 9.2 and the paper by K\'{o}mar et al. 2009). As a summary, we have that the number of oscillations
of orbital elements is 10/6 times higher than for the standard model of the galactic tide (see Figs. 2 and 7 in
K\'{o}mar et al. 2009). On the basis of Sec. 9.2 we know that the number of comets coming 
to distances less than 5 AU is 28.6/5.0 times higher than it is in the standard model and the frequency of
the returns may be even in 15/10 times greater (see Fig. 11). Thus we obtain that the number of occurence
of the long-period comets in distances less than 5 AU can be 15-times (10/6 $\times$ 15/10 $\times$ 28.6/5.0 $\doteq$ 15) 
greater than the standard model offers (see, e.g., Dones et al. 2004, pp. 161-162).

\subsubsection{The second access}
On the basis of Eqs. (\ref{eq:apnewf1}) and Eqs. (\ref{eq:dfaaad}) we obtain that the ratio between the frequencies of
cometary returns into the inner part of the Solar System for $\alpha$ $=$ $-$ 3.5, $a_{min}$ / $a_{max}$ $=$ 1/10
(Dones et al. 2004, Fern\'{a}ndez and Gallagher 1999) and $\alpha$ $=$ $-$ 0.55, $a_{min}$ $\rightarrow$ 0
(see Eqs. \ref{eq:nca2-jn}) is 
$f (0.55) $ / $f(3.5)$ $=$ ($\langle a \rangle _{3.5}$ / $\langle a \rangle _{0.55}$)$^{3}$ $=$ 7.
 On the basis of Sec. 9.2 we know that the number of comets coming 
to distances less than 5 AU is 28.6/5.0 times higher than it is in the standard model and the frequency of
the returns may be even in 15/10 times greater (see Fig. 11). Thus we obtain that the number of occurence
of the long-period comets in distances less than 5 AU can be 60-times (7 $\times$ 15/10 $\times$ 28.6/5.0 $\doteq$ 60) 
greater than the standard model offers (see, e.g., Dones et al. 2004, pp. 161-162).

\subsubsection{Discussion}
The two results were obtained, partially, in two different ways. The first case has not considered
any density function in semi-major axis $a$. However, it was based on numerical calculations presented by
K\'{o}mar et al. (2009) for $a$ $=$ 5 $\times$ 10$^{4}$ AU, while more correct access is to use the
mean value $\langle a \rangle$ as it is in the second case. In any case, both accesses yield values which are
more than 10-times smaller than the conventional values (see, e.g., Dones et al. 2004).

\subsection{Our estimate of the mass of the Oort cloud}
On the basis of the results discussed above we can come to the conclusion that the mass of the Oort cloud 
is less than 0.5 mass of the Earth (maybe, even 0.1 $M_{E}$). However, inclusion of perturbation by close
stars and interstellar clouds may change the result.

We can present more exact calculation based on the above presented values.
The current estimate of the mass of the Oort cloud of comets is (3.3 $-$ 7.0) masses of the Earth ($M_{E}$),
(Dones et al. 2004, p. 162), or $M'$ $=$ ( 5.1 $\pm$ 1.9 ) $M_{E}$. Our two, partially independent, results
yield that the real mass $M$ is (15 $-$ 60)$-$times smaller than the value $M'$, or,  $M$ is $d$ $=$ (37 $\pm$ 23)$-$times 
smaller than $M'$. Thus, $M$ $=$ $M'$ / $d$. Using also error analysis, we finally obtain
$M$ $=$ ( 0.14 $\pm$ 0.10 ) $M_{E}$. 

We have not considered the effect of nearby stars (interstellar clouds). 
Let the stars can generate $N-$times higher number of observable comets than the conventional/standard model of
galactic tides offers. The real number of observable comets, due to the action of more realistic galactic tides, is
[ (15-60) $+$ $N$ ] / ( 1 $+$ $N$  ) greater than the conventional estimate. Thus, the mass of the Oort cloud is
( 1 $+$ $N$ ) / [ (15-60) $+$ $N$ ] lower than  the conventional estimate. The formula yields, as an example:
$M$ $\approx$ (1/20) $M'$ $\approx$ (1/4) $M_{E}$ for $N$ $=$ 1, 
$M$  $\approx$ (1/13) $M'$ $\approx$ (1/3) $M_{E}$ for $N$ $=$ 2, 
$M$ $\approx$ (1/7) $M'$ $\approx$ (5/7) $M_{E}$ for $N$ $=$ 5, 
$M$ $\approx$ (1/5) $M'$ $\approx$ 1 $M_{E}$ for $N$ $=$ 10. 
If we take into account the result of Rickman et al. (2008, Fig. 2),
then we should use, as an approximation, $N$ $=$ 1 (we remind that the authors
use $\alpha$ $=$ $-$1.5 which significantly differs from 0).

\section{Summary}
The paper presents various results on the Oort cloud of comets if gravity of the Sun, Galaxy
(and Jupiter in Sec. 9) are considered.

Sec. 2 discusses the (density) function of inclination. Besides the first theoretical part,
the dominant part deals with the inclination with respect to the galactic equatorial plane. 
The important result shows that the exponent $\alpha$, characterizing distribution of comets 
in the Oort cloud as a function of semi-major axis, should be equal to $-$ 1 if $a_{min}$ 
corresponds to (10 $-$ 20) $\times$ 10$^{3}$ AU. This is not consistent with the
cases treated in the literature (Duncan et al. 1987, Fern\'{a}ndez and Ip 1987, Fern\'{a}ndez 1992, 
Fern\'{a}ndez and Gallardo 1999). If $a_{min}$ $\ll$ (10 $-$ 20) $\times$ 10$^{3}$ AU, 
then various values of $\alpha$ are admitted in our treatment of the distribution in inclinations.

Secs. 3 and 4 present simple relations for some dependencies. They both improve the published results
and found new results. The relation between the semi-major axis $a$ and oscillation period $P$ is some kind
of analogy to the third Kepler's law. The relation reads $a^{3}$ $P$ $=$ 1 if $a$ and $P$ are measured in natural units.

Sec. 6 seems to be of theoretical character. However, its results are applied to practical problems.
One question remains open: Why various authors consider distribution in eccentricity in the form treated
by Jeans (1919) as the relevant theoretical access but the other part of the Jeans distribution, corresponding
to the density function for semi-major axis, is ignored? 
Results of Sec. 6 represented by Eqs. (\ref{eq:dfaq7})-(\ref{eq:dfaq8}) are improvements of the
published results (see e.g., Hills 1981, Fern\'{a}ndez and Gallardo 1999). The result represented by
Eq. (\ref{eq:dfaaa}) are used in Sec. 10.

Sec. 6 presents the effect of galactic tide to some observational quantities.
The cumulative number of comets with perihelion distances is described by Eqs. (\ref{eq:peri6}) and (\ref{eq:peri6-j}), 
which can be generalized to the form $N(<q)$ $=$ $N(<q_{0})$ $\left ( q / q_{0} \right ) ^{2/3}$, $q$, $q_{0}$ $\in$ ( 0, 100 ) AU. 
This is equivalent to the distribution function
\begin{eqnarray}\label {eq:peri6d}
F_{q} (q) &=&  \left ( \frac{q}{q_{0}} \right ) ^{2/3} ~, ~~ q_{0} = 100 ~
\mbox{AU}, ~~ q \in ( 0, q_{0} ) ~.
\end{eqnarray}
This result differs from the conventional result based on the Jeans density function $f(e)$ $=$ 2 $e$.
However, if the authors do not agree with the Jeans density function of semi-major axis
$f_{a} (a)$ $\propto$ $\sqrt{a}$, then they cannot use $f(e)$ $=$ 2 $e$. Correspondingly, 
the distribution function cannot be of the form 
$F_{q} (q)$ $=$ ($q$ / $a$) ($2~-~ q / a$).
 as the authors state (see the second of Eqs. \ref{eq:jeansfq3}).
 More correctly, if one wants to use  $f(e)$ $=$ 2 $e$, he should be able to present an argument
 in favor of the choice and the argument must be independent of the argument presented by Jeans (1919).
Moreover, the results of Sec. 6.4.3 show that $F_{q} (q)$ $\propto$ $q$ for the inner part of the Solar System
holds for any density function of the form $f(e)$ $=$ ( $k$ $+$ 1 ) $e^{k}$, $k$ $>$ 0, not only for $k$ $=$ 1.

Sec. 9 considers also the gravity of Jupiter. 
The gravitational action of Jupiter significantly influences the distribution in
perihelion distance. The real number of long-period comets may be about
50-times smaller than the number conventionally considered, as it is discussed in Sec. 10.

We have already mentioned the problem with the values of the exponent $\alpha$. The exponent 
characterizes the distribution of comets in the Oort cloud as a function of semi-major axis. 
We have found, in Sec. 2 for distribution in inclinations, that only the condition 
$a_{min}$ $\ll$ (10 $-$ 20) $\times$ 10$^{3}$ AU enables $\alpha$ $\ne$ $-$ 1.
Sec. 10 deals with the value of the $\alpha$ in a different way. Sec. 10 comes to the
conclusion that $\alpha$ $=$ $-$ 0.55 and $a_{min}$ $\rightarrow$ 0, see Eqs. (\ref{eq:nca2-jn}).
Secs. 6.3 and 9.2 yield $F_{q} (q)$ $\propto$ $q^{2/3}$ if also $q$ greater than several astronomical units are considered.
If this holds also for $q$ $>$ $a_{min}$, then Eqs. (\ref{eq:dfaq8}) yields $\alpha$ $=$ $-$ 1 / 3.  This is, approximately,
consistent with the result of Eq. (\ref{eq:nca2-jn}),  $\alpha$ $=$ $-$ 0.55. Thus, we can conclude that 
$\alpha$ is about $-$ 1/2, more correctly, $\alpha$ $\in$  ( $-$0.6, $-$ 0.3 ). 

Consideration of the results of the previous sections yields that mass of the Oort cloud of comets
is ( 0.14 $\pm$ 0.10 ) masses of the Earth if the action of the nearby stars (galactic clouds) is
negligible. The mass of the Oort cloud is less than 1 $M_{E}$ even if the effect of the nearby stars is important.

\section{Conclusion}
The main results are: \\
1. Theoretical number of long-period comets with perihelion distance $q$ $<$ 5 AU is about 50-times
greater than the conventional approach yields. (Gravity of Jupiter was taken into account in finding this result.) 
Mass of the Oort cloud of comets is, probably, about 1/4  mass of the Earth. \\
2. Semi-major axis $a$ and period of oscillations $P$ of eccentricity
(and other orbital elements) are related as $a^{3}$ $P$ $=$ 1 
in natural units for a moving Solar System in the Galaxy. 
The natural unit for time is the orbital period of the Solar System revolution around the galactic center
and the natural unit for measuring the semi-major axis is its maximum value for the half-radius of the
Solar System corresponding to the half-radius of the Oort cloud. 
The relation holds for the cases when comets approach the inner part of the Solar System,
e.g., perihelion distances are less than $\approx$ 100 AU. \\
3. The minimum value of semi-major axis for the Oort cloud is 
$a_{min}$ $\ll$ 1 $\times$ 10$^{4}$ AU. This condition was obtained both from the numerical 
results on cometary evolution under the action of the galactic tides and from the observational
distribution of long-period comets. If the density function of semi-major axis is approximated by 
proportionality $a^{\alpha}$, then $\alpha$ is, approximately, $-$ 1/2. \\
4. The magnitude of the change in perihelion distance per orbit, $\Delta q$, of a comet due to galactic tides 
 is a strong function of semi-major axis $a$, proportional to $a^{8.25}$. 

The usage of the density function of eccentricity in the form  $f(e)$ $=$ 2 $e$ cannot be 
argumented to be the result obtained by Jeans (1919). Consequence of the Jeans calculations is marginal 
density function of semi-major axis proportional to $\sqrt{a}$ and this is not consistent with the conventionally
used distributions. Moreover, Figs. 7 and 8 in K\'{o}mar et al. (2009) show that
gravitational tides alone would produce only values $e$ $<$ 0.9 for galactic inclinations $i$
less than 90 degrees and practically $e$ $<$ 0.8 for $i$ $\approx$ 90 degrees: only for very short time
 intervals $e$ is greater than 0.8 ($e$ is close to 1 during the cometary visit of the inner part of the Solar System).

Using the observations yielding the distribution function of perihelion distance $F_{q} (q)$ $\propto$ $q$ (inner part of the Solar System),
we cannot come to the conclusion that  $f(e)$ $=$ 2 $e$. Any density function of the form $f(e)$ $=$ ( $k$ $+$ 1 ) $e^{k}$, $k$ $>$ 0
is also consistent with the observational result $F_{q} (q)$ $\propto$ $q$. 

Observational data on distribution in semi-major axis $a$, theoretical analytical results on the distribution functions $F_{a} (a)$ 
and of a perihelion distance, and, detailed numerical calculations on gravitational action of the Galaxy lead to a conclusion on $F_{a} (a)$.
The conclusion is that approximation of $F_{a} (a)$ by the form proportional to $a^{\alpha + 1}$ is consistent with all the mentioned
methods if $\alpha$ is about $-$ 1/2.

The obtained results may be improved to be more accurate. This will require also more robust 
and more detailed numerical calculations (and, also, several planets, not only Jupiter, have to be considered).

\section*{Appendix A: Inclination with respect to the ecliptical plane}

(Reference to equation of number (j) of this appendix is denoted as Eq. (A~j).
Reference to equation of number (i) of the main text is denoted as Eq. (i).)

\setcounter{equation}{0}

We are interested in the inclination with respect to the ecliptical plane
if we know inclination (and longitude of the ascending node) with respect
to the galactic equatorial plane.

The transformation will be found in several steps.

We will use $x_{i}$, $i$ $=$ 1, 2, 3 (and also primed quantities
for other reference frames, all with origin in the Sun)
for coordinate right-handed axes. Summation convention is adopted, i.e. summation
over repeated indices is assumed.

At first, we will find unit vector normal to the (osculating) orbital plane
of the comet. Let the orbital plane is characterized with the
ascending node $\Omega$ and galactic inclination $i$ in the unprimed system.
In order to obtain coordinates in the system where preferred plane is
given by the orbital plane of the comet, we will make two transformations:
\begin{eqnarray}\label{A1}
x_{i} ' &=& A_{ij} ~x_{j} ~,
\nonumber \\
A_{11} &=& +~ \cos \Omega ~, ~~
A_{12} = +~ \sin \Omega ~, ~~
A_{13} = 0 ~,
\nonumber \\
A_{21} &=& -~ \sin \Omega ~, ~~
A_{22} = +~ \cos \Omega ~, ~~
A_{23} = 0 ~,
\nonumber \\
A_{31} &=& 0 ~, ~~
A_{32} = 0 ~, ~~
A_{33} = +~ 1 ~,
\end{eqnarray}
and,
\begin{eqnarray}\label{A2}
x_{i} '' &=& B_{ij} ~x_{j} ' ~,
\nonumber \\
B_{11} &=& +~1 ~, ~~
B_{12} = 0 ~, ~~
B_{13} = 0 ~,
\nonumber \\
B_{21} &=& 0~, ~~
B_{22} = +~ \cos i ~, ~~
B_{23} =  +~ \sin i ~,
\nonumber \\
B_{31} &=& 0 ~, ~~
B_{32} = -~ \sin i ~, ~~
B_{33} = +~ \cos i ~.
\end{eqnarray}
Eqs. (A1) and (A2) yield ($C_{ij}$ $=$ $B_{ik}$ $A_{kj}$)
\begin{eqnarray}\label{A3}
x_{i} '' &=& C_{ij} ~x_{j}  ~,
\nonumber \\
C_{11} &=& +~ \cos \Omega ~, ~~
C_{12} = +~ \sin \Omega ~, ~~
C_{13} = 0 ~,
\nonumber \\
C_{21} &=& -~ \sin \Omega ~ \cos i ~, ~~
C_{22} = +~ \cos \Omega ~ \cos i ~, ~~
C_{23} =  +~ \sin i ~,
\nonumber \\
C_{31} &=& +~ \sin \Omega ~ \sin i ~, ~~
C_{32} = -~ \cos \Omega ~ \sin i ~, ~~
C_{33} = +~ \cos i ~.
\end{eqnarray}
The inverse transformation is (transformation matrix $C^{T}$ is inverse,
in our case transpose, to the matrix $C$):
\begin{eqnarray}\label{A4}
x_{i}  &=& C_{ij}^{T} ~x_{j} ''  ~,
\nonumber \\
C_{11}^{T} &=& +~ \cos \Omega ~, ~~
C_{12}^{T} = -~ \sin \Omega ~ \cos i  ~, ~~
C_{13}^{T} = +~ \sin \Omega ~ \sin i  ~,
\nonumber \\
C_{21}^{T} &=& +~ \sin \Omega ~, ~~
C_{22}^{T} = +~ \cos \Omega ~ \cos i ~, ~~
C_{23}^{T} = -~ \cos \Omega ~ \sin i ~,
\nonumber \\
C_{31}^{T} &=& 0 ~, ~~
C_{32}^{T} =  +~ \sin i ~, ~~
C_{33}^{T} = +~ \cos i ~.
\end{eqnarray}

The unit vector normal to the orbital plane of the comet is characterized
by the condition
\begin{eqnarray}\label{A5}
( x_{1} '', x_{2} '', x_{3} '' )^{T} &=& (0, 0, 1)^{T} ~.
\end{eqnarray}
Its galactic coordinates are, using Eqs. (A4),
\begin{eqnarray}\label{A6}
\vec{e}_{N} &\equiv& ( x_{1}, x_{2}, x_{3} )^{T} ~,
\nonumber \\
x_{1} &=& +~ \sin \Omega ~ \sin i ~,
\nonumber \\
x_{2} &=& -~ \cos \Omega ~ \sin i ~,
\nonumber \\
x_{3} &=&  +~ \cos i ~.
\end{eqnarray}

The top of the vector $\vec{e}_{N}$ is a point. The point has the following
galactic coordinates
\begin{eqnarray}\label{A7}
\vec{e}_{N} &\equiv& ( x_{1}, x_{2}, x_{3} )^{T} ~,
\nonumber \\
x_{1} &=& +~ \cos l ~ \cos b ~,
\nonumber \\
x_{2} &=& +~ \sin l ~ \cos b ~,
\nonumber \\
x_{3} &=&  +~ \sin b ~,
\end{eqnarray}
where $l$ is the galactic longitude and $b$ is the galactic latitude.

Comparison of Eqs. (A6)-(A7) yields
\begin{eqnarray}\label{A8}
\cos l ~ \cos b &=& +~ \sin \Omega ~ \sin i ~,
\nonumber \\
\sin l ~ \cos b  &=& -~ \cos \Omega ~ \sin i ~,
\nonumber \\
\sin b	&=&  +~ \cos i ~.
\end{eqnarray}
If $\Omega$ and $i$ are given, then Eqs. (A8) offer
\begin{eqnarray}\label{A9}
b  &=&	\arcsin ( \cos i ) ~,
\nonumber \\
\cos l	&=& +~ \frac{1}{\cos b} ~ \sin \Omega ~ \sin i ~,
\nonumber \\
\sin l &=& -~ \frac{1}{\cos b} ~ \cos \Omega ~ \sin i ~.
\nonumber \\
\end{eqnarray}
We can easily find $l$ from Eqs. (A9). It is sufficient to use the following
prescription:
\begin{eqnarray}\label{A10}
\cos \Psi  &\equiv& C ~,
\nonumber \\
\sin \Psi &\equiv& S ~,
\nonumber \\
S \ge 0 &\Rightarrow& \Psi = \arccos ~ C ~,
\nonumber \\
S \le 0 &\Rightarrow& \Psi = 2 ~\pi ~-~ \arccos ~C ~.
\end{eqnarray}

We can summarize: We have $i$ and $\Omega$ of a cometary orbit with respect
to the galactic equatorial plane. The corresponding unit vector normal to the
cometary orbit $\vec{e}_{N}$ is given by Eqs. (A6).
Galactic coordinates $l$ and $b$ of the unit vector are given by
Eqs. (A7) and they can be found using Eqs. (A8)-(A10).

Equatorial coordinates of the vector $\vec{e}_{N}$ are characterized by the
right ascension $\alpha$ and the declination $\delta$:
($\vec{e}_{N~eq.c.}$)$_{1}$ $=$ $\cos \alpha ~ \cos \delta$,
($\vec{e}_{N~eq.c.}$)$_{2}$ $=$ $\sin \alpha ~ \cos \delta$,
($\vec{e}_{N~eq.c.}$)$_{3}$ $=$ $\sin \delta$.
Transformations between galactic and equatorial coordinates yield
\begin{eqnarray}\label{A11}
\sin \delta  &=& \sin b ~ \sin ~\delta_{0} ~+~
\cos b ~ \cos ~\delta_{0} ~\sin ( l - l_{0} ) ~,
\nonumber \\
\cos ( \alpha - \alpha_{0} ) ~ \cos \delta  &=& \sin b ~ \cos ~\delta_{0} ~-~
\cos b ~ \sin ~\delta_{0} ~\sin ( l - l_{0} ) ~,
\nonumber \\
\sin ( \alpha - \alpha_{0} ) ~ \cos \delta  &=&
\cos b ~ ~\cos ( l - l_{0} ) ~,
\nonumber \\
\alpha_{0} &=& 192.86 ^{\circ} ~,
\nonumber \\
\delta_{0} &=& 27.13 ^{\circ} ~,
\nonumber \\
l_{0} &=& 33.93 ^{\circ} ~.
\end{eqnarray}

If $\Omega$ and $i$ are given, then Eqs. (A8)-(A11) enable to find
$\alpha$ and $\delta$. If $i$ $=$ 0, then $\delta$ $=$ $\delta_{0}$ and
$\alpha$ $=$ $\alpha_{0}$.

Ecliptical coordinates of the vector $\vec{e}_{N}$ are characterized by
the ecliptical longitude $\lambda$ and the ecliptical latitude $\beta$:
($\vec{e}_{N~ecl.c.}$)$_{1}$ $=$ $\cos \lambda ~ \cos \beta$,
($\vec{e}_{N~ecl.c.}$)$_{2}$ $=$ $\sin \lambda ~ \cos \beta$,
($\vec{e}_{N~ecl.c.}$)$_{3}$ $=$ $\sin \beta$.
Transformations between equatorial and ecliptical coordinates yield
\begin{eqnarray}\label{A12}
\sin \beta  &=& \sin \delta ~ \cos ~\varepsilon ~-~
\cos \delta ~ \sin ~\varepsilon ~\sin \alpha ~,
\nonumber \\
\cos \lambda ~ \cos \beta  &=& \cos \delta ~ \cos \alpha ~,
\nonumber \\
\sin \lambda ~ \cos \beta  &=&
\sin \delta ~\sin \varepsilon ~+~ \cos \delta ~\cos \varepsilon ~ \sin \alpha ~,
\nonumber \\
\varepsilon &=& 23.5 ^{\circ} ~.
\end{eqnarray}

If $\Omega$ and $i$ are given, then Eqs. (A8)-(A11) enable to find
$\alpha$ and $\delta$. If $i$ $=$ 0, then $\delta$ $=$ $\delta_{0}$ and
$\alpha$ $=$ $\alpha_{0}$.
If $\alpha$ and $\delta$ are known, then Eqs. (A10) and (A12) enable to find
$\lambda$ and $\beta$. If $\delta$ $=$ 90$^{\circ}$, then
$\lambda$ $=$ 90$^{\circ}$ and
$\beta$ $=$ $\arcsin ( \cos \varepsilon )$.

Finally, the longitude of the ascending node $\Omega_{ecl}$ and the inclination
$i_{ecl}$ of the cometary orbit measured in the ecliptical coordinate system
are given as follows:
\begin{eqnarray}\label{A13}
+~ \sin \Omega_{ecl} ~\sin i_{ecl} &=& \cos \lambda ~ \cos \beta ,
\nonumber \\
-~ \cos \Omega_{ecl} ~ \sin i_{ecl} &=& \sin \lambda ~ \cos \beta  ~,
\nonumber \\
+~ \cos i_{ecl} &=& \sin \beta	~.
\end{eqnarray}
The solution of Eqs. (A13) is
\begin{eqnarray}\label{A14}
i_{ecl} &=& \arccos ( \sin \beta )  ~,
\nonumber \\
\sin \Omega_{ecl} &=& +~ \frac{1}{\sin i_{ecl}} ~\cos \lambda ~ \cos \beta ,
\nonumber \\
\cos \Omega_{ecl} &=& -~ \frac{1}{\sin i_{ecl}} ~\sin \lambda ~ \cos \beta  ~,
\end{eqnarray}
where also Eqs. (A10) must be used.

\section*{Appendix B: Motion of Jupiter in the ecliptical plane}

(Reference to equation of number (j) of this appendix is denoted as Eq. (B~j).
Reference to equation of number (i) of the main text is denoted as Eq. (i).)

\setcounter{equation}{0}

If we want to take into account also gravity of a planet, we need
to find its position in the galactic coordinates. Let us consider
Jupiter moving in a circular orbit in the ecliptical plane. 
Using results of the Appendix A, we can write for Jupiter coordinates
\begin{eqnarray}\label{B1}
x_{1~J} '' &=& a_{J} ~ \cos ( \omega_{J} ~t ~+~ \varphi_{0} ) ~,
\nonumber \\
x_{2~J} '' &=& a_{J} ~ \sin ( \omega_{J} ~t ~+~ \varphi_{0} ) ~,
\nonumber \\
x_{3~J} '' &=& 0 ~,
\nonumber \\
\omega_{J}  &=& \frac{2 ~\pi}{a_{J}^{3/2}} ~yr^{-1}~,
\nonumber \\
a_{J} &=& 5.203 ~ \mbox{AU} ~,
\end{eqnarray}
where $t$ is the time and $\varphi_{0}$ is the arbitrary initial phase.
The unit vector normal to the orbital plane of the comet is characterized
by the condition
\begin{eqnarray}\label{B2}
( x_{1~J} '', x_{2~J} '', x_{3~J} '' )^{T} &=& (0, 0, 1)^{T} ~.
\end{eqnarray}
As for the galactic coordinates, we have (see Eqs. A6)
\begin{eqnarray}\label{B3}
( \vec{e}_{N} )_{gal} &\equiv& ( x_{1}, x_{2}, x_{3} )^{T} ~,
\nonumber \\
x_{1} &=& +~ \sin \Omega_{J} ~ \sin i_{J}  ~,
\nonumber \\
x_{2} &=& -~ \cos \Omega_{J}  ~ \sin i_{J}  ~,
\nonumber \\
x_{3} &=&  +~ \cos i_{J}  ~.
\end{eqnarray}

Eqs. (A4) yield
\begin{eqnarray}\label{B4}
x_{1~J} &=& x_{1~J} '' ~\cos \Omega_{J} ~-~
x_{2~J} '' ~ \sin \Omega_{J} ~ \cos i_{J} ~,
\nonumber \\
x_{2~J} &=& x_{1~J} '' ~\sin \Omega_{J} ~+~
x_{2~J} '' ~ \cos \Omega_{J} ~ \cos i_{J} ~,
\nonumber \\
x_{3~J} &=& x_{2~J} '' ~\sin i_{J} ~.
\end{eqnarray}

We need the values of $\Omega_{J}$ and $i_{J}$.

We have ($\vec{e}_{N}$)$_{ecl}$ $=$ ( 0, 0, 1 )$^{T}$.
Transformations inverse to Eqs. (A12) 
\begin{eqnarray}\label{B5}
\sin \delta  &=& \sin \beta ~ \cos ~\varepsilon ~+~
\cos \beta ~ \sin ~\varepsilon ~\sin \lambda ~,
\nonumber \\
\cos \alpha ~ \cos \delta  &=& \cos \beta ~ \cos \lambda ~,
\nonumber \\
\sin \alpha ~ \cos \delta  &=& -~
\sin \beta ~\sin \varepsilon ~+~ \cos \beta ~\cos \varepsilon ~ \sin \lambda ~,
\nonumber \\
\varepsilon &=& 23.5 ^{\circ} ~,
\end{eqnarray}
yield for $\beta$ $=$ 90$^{\circ}$
\begin{eqnarray}\label{B6}
\sin \delta_{J}  &=& +~ \cos ~\varepsilon  ~,
\nonumber \\
\cos \alpha_{J} ~ \cos \delta_{J}  &=& 0 ~,
\nonumber \\
\sin \alpha_{J} ~ \cos \delta_{J}  &=& -~ \sin \varepsilon  ~,
\nonumber \\
\varepsilon &=& 23.5 ^{\circ} ~.
\end{eqnarray}
Solution of Eqs. (B6) is
\begin{eqnarray}\label{B7}
\delta_{J}  &=& \arcsin ( \cos ~\varepsilon ) ~,
\nonumber \\
\alpha_{J} &=& 270  ^{\circ} ~,
\nonumber \\
\varepsilon &=& 23.5 ^{\circ} ~.
\end{eqnarray}
Thus, we have
\begin{eqnarray}\label{B8}
(\vec{e}_{N})_{eq} &=& ( \cos \alpha_{J} ~ \cos \delta_{J}, \sin \alpha_{J} ~ \cos \delta_{J}, \sin \delta_{J} )^{T}
\nonumber \\
&=& (0, -~ \sin \varepsilon,  +~ \cos \varepsilon )^{T} ~, 
\nonumber \\
\varepsilon &=& 23.5 ^{\circ} ~.
\end{eqnarray}

As for transformation to the galactic coordinate system, we have to use transformations inverse to those
represented by Eqs. (A11):
\begin{eqnarray}\label{B9}
\sin b  &=& \sin \delta ~ \sin ~\delta_{0} ~+~
\cos \delta ~ \cos ~\delta_{0} ~\cos ( \alpha - \alpha_{0} ) ~,
\nonumber \\
\cos ( l - l_{0} ) ~ \cos b  &=& \cos \delta ~ \sin ( \alpha - \alpha_{0} ) ~,
\nonumber \\
\sin ( l - l_{0} ) ~ \cos b  &=& \sin \delta ~\cos ~\delta_{0} ~-~ 
\cos ~\delta ~ \sin ~\delta_{0} ~ \cos (  \alpha - \alpha_{0} ) ~,
\nonumber \\
\alpha_{0} &=& 192.86 ^{\circ} ~,
\nonumber \\
\delta_{0} &=& 27.13 ^{\circ} ~,
\nonumber \\
l_{0} &=& 33.93 ^{\circ} ~,
\end{eqnarray}
which yield, together with Eqs. (B7), 
\begin{eqnarray}\label{B10}
\sin b_{J}  &=&  \sin ~\delta_{0} ~ \cos \varepsilon ~-~
\cos ~\delta_{0} ~\sin \varepsilon ~\sin  \alpha_{0}  ~,
\nonumber \\
\cos ( l_{J} - l_{0} ) ~ \cos b_{J}  &=& -~ \cos \alpha_{0}  ~ \sin \varepsilon ~,
\nonumber \\
\sin ( l_{J} - l_{0} ) ~ \cos b_{J}  &=& +~ \cos ~\delta_{0} ~ \cos \varepsilon ~+~ 
\sin ~\delta_{0} ~ \sin \varepsilon ~ \sin \alpha_{0}  ~.
\end{eqnarray}
Finally, we can write
\begin{eqnarray}\label{B11}
(\vec{e}_{N})_{gal} &=& ( \cos l_{J} ~ \cos b_{J}, \sin l_{J} ~ \cos b_{J}, \sin b_{J} )^{T}
\nonumber \\.
&=& ( \sin \Omega_{J} ~  \sin i_{J} , -~  \cos \Omega_{J} ~ \sin i_{J} ,  \cos i_{J}  )^{T} ~. 
\end{eqnarray}
Eqs. (B10)-(B11) give
\begin{eqnarray}\label{B12}
i_{J}  &=&  \arccos \left \{ \sin ~\delta_{0} ~ \cos \varepsilon ~-~
\cos ~\delta_{0} ~\sin \varepsilon ~\sin  \alpha_{0} \right \}   ~,
\nonumber \\
\cos b_{J} &=& \left \{ 1 ~-~ \left (   \sin ~\delta_{0} ~ \cos \varepsilon ~-~
\cos ~\delta_{0} ~\sin \varepsilon ~\sin  \alpha_{0} \right )^{2}  \right \}^{1/2}
\nonumber \\
\cos ( l_{J} - l_{0} ) &=& \frac{1}{\cos b_{J}} ~\left (  -~ \cos \alpha_{0}  ~ \sin \varepsilon \right )  ~,
\nonumber \\
\sin ( l_{J} - l_{0} )  &=& \frac{1}{\cos b_{J}} ~\left (  \cos ~\delta_{0} ~ \cos \varepsilon ~+~ 
\sin ~\delta_{0} ~ \sin \varepsilon ~ \sin \alpha_{0} \right )   ~,
\nonumber \\
\sin \Omega_{J} &=& +~ \frac{1}{\sin i_{J}} ~ \cos l_{J} ~ \cos b_{J} ~,
\nonumber \\
\cos \Omega_{J} &=& -~ \frac{1}{\sin i_{J}} ~ \sin l_{J} ~ \cos b_{J} ~.
\end{eqnarray}
Eqs. (A10) must be used in finding $l_{J}$ and $\Omega_{J}$.

\section*{Acknowledgement}
This work was supported by the Scientific Grant Agency VEGA,
Slovak Republic, grant No. 2/0016/09.

\end{document}